\def\lesssim{{_ <\atop{^\sim}}}
\def\grtsim{{_ >\atop{^\sim}}}
\def\ap3m{AP$^3$M}
\def\hkpc{h^{-1}{\ }{\rm kpc}}
\def\hMpc{h^{-1}{\ }{\rm Mpc}}
\def\hMsun{h^{-1}{\ }{\rm M_{\odot}}}
\def\kms{{\rm{\ }km{\ }s^{-1}}}

\def\profiles{4.5}

\def\eq_s2{(7)}

\documentstyle[epsfig]{mn}

\begin{document}

   \title{On effects of resolution in dissipationless cosmological simulations}

   \author[Knebe et al.]
          {Alexander Knebe,$^{1,2}$
           Andrey V. Kravtsov,$^{3,4}$
          Stefan Gottl\"ober$^1$
          \newauthor
          and
          Anatoly A. Klypin$^3$\\
          $^1$ Astrophysikalisches Institut Potsdam (AIP), 
               An der Sternwarte 16, 14482 Potsdam, Germany\\
          $^2$ Theoretical Physics, 1 Keble Road, Oxford OX1 3NP, UK\\
          $^3$ Astronomy Department, New Mexico State University,
               Dept.4500, Las Cruces, NM 88003-0001, USA\\
          $^4$ Current address: Department of Astronomy, The Ohio State University, 140 W. 18th Ave., Columbus, OH 43210-1173, USA
          }

   \date{Received ...; accepted ...}

   \maketitle

\begin{abstract}    
 We present a study of numerical effects in dissipationless
 cosmological simulations. The numerical effects are evaluated and
 studied by comparing results of a series of $64^3$-particle
 simulations of varying force resolution and number of time steps,
 performed using three of the $N$-body techniques currently used for
 cosmological simulations: the Particle Mesh (PM), the Adaptive
 Particle-Particle Particle-Mesh (\ap3m), and the newer Adaptive
 Refinement Tree (ART) code. This study  can therefore be interesting
 both as an analysis of numerical effects and as a systematic comparison
 of different codes. 

We find that the \ap3m and the ART codes produce similar results,
given that convergence is reached within the code type. We also
find that numerical effects may affect the high-resolution simulations
in ways that have not been discussed before. In particular, our study
revealed the presence of two-body scattering, effects of which can
be greatly amplified by inaccuracies of time integration. This process
appears to affect the correlation function of matter, mass function
and inner density of dark matter halos and other statistics at scales
much larger than the force resolution, although different statistics
may be affected in different fashion. We discuss the conditions at
which strong two-body scattering is possible and discuss
the choice of the force resolution and integration time
step. Furthermore, we discuss recent claims that simulations with
force softening smaller than the mean interparticle separation are
not trustworthy and argue that this claim is incorrect in general
and applies only to the phase-sensitive statistics. Our conclusion is
that, depending on the choice of mass and force resolution and
integration time step, a force resolution as small as 
$0.01$ of the mean interparticle separation can be justified.
\end{abstract}

\begin{keywords}
cosmology -- numerical simulations.
\end{keywords}

\section{Introduction}

Dissipationless cosmological $N$-body simulations are currently the tool
of choice for following the evolution of cold dark matter (CDM) into the highly
nonlinear regime. For the widest range of plausible dark matter (DM)
candidates (from axions of mass $\sim 10^{-6}-10^{-3}{\ }{\rm eV}$ to
$\sim 10^{21}-10^{25}{\ }{\rm eV}$ wimpzillas; see, e.g., Kolb, Chung
\& Riotto 1998; Roszkowski 1999), their expected number density is 
$n_{\rm DM}\sim (10^{52}-10^{83})\Omega_0 h^2{\ }{\rm Mpc^{-3}}.$
$N$-body simulations numerically solve the $N$-body problem: given
initial positions and velocities for $N$ pointlike massive objects,
the simulations predict the particle positions and velocities at any
subsequent time. Current $N$-body simulations are capable of following
the evolution of $\lesssim 10^9$ particles, far short of the expected
number of DM particles.  Therefore, the correct approach to modelling
dark matter evolution in a cosmologically representative volume, is to
use the Vlasov equation (collisionless Boltzmann equation) coupled
with the Poisson equation and complemented by appropriate boundary
conditions.  However, a full-scale modelling of 6D distribution
functions with  reasonable spatial resolution is extremely
challenging computationally. The alternative approach adopted in most
cosmological simulations is to split the initial phase space
into small volume elements and follow evolution of these elements
using $N$-body techniques. Each volume element can thus be thought of
as an $N$-body particle, which moves with a flow and which has some
shape (for example, a box or a sphere) or is simply point-like. Two
particles are assumed to interact gravitationally only if they are
separated by a distance $\grtsim \epsilon$, where $\epsilon$ is the
{\em smoothing scale}, often referred to as the {\em force
resolution}.  It is clear that the ``size'' of a particle is
determined by the eulerian spatial size of the initial phase space
volume element or by the smoothing scale, whichever is smaller.

While this approach seems logical and reasonable and is expected to
provide approximate solution to a complicated problem, questions
of its limitations may be raised. For example, while particle shape is
usually considered to be rigid (fixed by a specific form of the Green
function or the shape of the interparticle force), in eulerian space
the shape of the initially cubic phase space volume element can be
expected to be stretched as the element moves towards higher density
regions. Its volume can also change so as to preserve the phase space
density.  Furthermore, under certain conditions the $N$-body systems
may exhibit scattering, which is undesirable when one models a purely
collisionless system. This may occur in cosmological simulations if the
``size'' of particles is much smaller than the {\em eulerian\/}
spatial size of the phase space element they are supposed to
represent.

These effects may influence the accuracy of the simulations and lead
to spurious results. Nevertheless, surprisingly little attention has been
given to studies of such limitations in the cosmological simulations.
As the resolution of simulations improves and the range of their
applications broaden, it becomes increasingly important to address
these issues.  Indeed, during the past decade the force resolution of
the simulations has improved by a factor of $\sim 100-1000$, while
(with rare exceptions) the mass resolution has improved only by a
factor of $\sim 10$. Also, modern high-resolution codes follow
evolution of cosmological systems for many dynamical time-scales. In
this regime the accuracy of the force estimates may be less important
than the stability of the overall solution. These issues are usually
not addressed when tests of codes are presented.

Recently, in a series of papers Melott and collaborators (Kuhlman,
Melott \& Shandarin 1996; Melott et al. 1997; Splinter et al. 1998)
raised the issue of the balance between the force and the mass
resolution.  While we disagree with their main conclusion that the
force resolution should be larger than the mean interparticle
separation (see \S~3 and \S~5), we agree that the issue is important.

There is also a common misconception related to the adaptive mesh
refinement approach in cosmological simulations and other algorithms
that integrate equations of particle motion in {\em comoving}
coordinates.  The common criticism (e.g., Splinter et al. 1998 and
references therein) is that these algorithms attempt to resolve scales
unresolved in the initial conditions (the scales below approximately
half of the Nyquist frequency). However, the goal of increasing the
resolution in comoving coordinates is not to resolve the waves not
present in the initial conditions but rather to properly follow all of
the waves initially present.

When followed in comoving coordinates, gravitational instability leads
to the separation of structures from the Hubble flow and collapse,
resulting in the transfer of power to higher wavenumbers. If the force
resolution is fixed in {\em comoving} coordinates at the Nyquist
frequency of the initial conditions, this transfer cannot be modelled
properly for all waves. Moreover, the size of structures that have
collapsed and virialized stays fixed in the proper coordinates and
decreases in comoving coordinates. When the comoving size of such
objects becomes smaller than the resolution of a fixed grid
simulation, their subsequent evolution and internal properties will be
modelled incorrectly. Adaptive mesh refinement algorithms address this
problem by increasing the force resolution locally to follow the
evolution of collapsing and virialized density peaks as their size
becomes less than the resolution of the original grid. Other
algorithms can achieve the same result by varying the comoving force
softening with time.

Our motivation for the present study is twofold. Our first goal is to
elaborate on the issue of spurious numerical effects. Namely, we study
the effects of the balance of force and mass resolutions and of time
integration details on statistics commonly used
in analyses of cosmological simulations. The balance of force and mass
resolutions should be studied by varying both of the resolutions. In
this study, however, we will keep the mass resolution fixed and vary
the force resolution instead. While this may not reveal all of the
artificial effects\footnote{For example, fixed mass resolution does
not allow us to study the effects of the rigid particle shape
mentioned above.}, this allows to study the spurious two-body
scattering present when the smoothing scale is set to be too small. It
also allows us to study how the limited force resolution affects
various statistical properties of the dark matter distribution.  Our
second goal is to compare results produced using two of the currently
employed high-resolution $N$-body codes:
\ap3m (Couchman 1991) and ART (Kravtsov, Klypin \& Khokhlov 1997). 
The comparison is of interest due to a relative novelty of the latter
technique and some disagreement between the results concerning the
details of the central density distribution in DM halos obtained using
different codes (e.g., Moore 1994; Navarro, Frenk \& White 1997;
Kravtsov et al. 1998; Moore et al. 1999). This study can also be
interesting as an independent test of the widely used \ap3m code, for
which virtually no systematic tests have been published to date.

The paper is organized as follows. In \S 2 we describe the $N$-body
algorithms used in our study and describe the numerical simulations
performed . In \S 3 we compare the dark matter distribution simulated
using different codes with different force resolutions and time
steps. In \S 4 we use these simulations to compare the properties and
distribution of dark matter halos (dense virialized systems). In \S 5
we summarize the results of the code comparisons, discuss the effects
of resolution and time step on the commonly used statistics, and
present our conclusions.

\section{Cosmological $N$-Body Simulations}

In this study we will use and compare three different $N$-body
algorithms: Particle-Mesh (PM) algorithm (Hockney \& Eastwood 1981),
adaptive Particle-Particle Particle-Mesh algorithm (\ap3m; Couchman
1991), and Adaptive Refinement Tree algorithm (ART; Kravtsov et
al. 1997). The PM algorithm was first used for cosmological
simulations by Doroshkevich et al. (1980), Efstathiou \& Eastwood
(1981), and Klypin \& Shandarin (1983). The algorithm makes use of
the fast Fourier transforms to solve the Poisson equation on a uniform
grid and uses interpolation and numerical differentiation to obtain
the force that acts on each particle. 

The solution is limited by the number of particles (mass resolution)
and by the size of the grid cell which defines force resolution. The
exact shape of the resulting force depends on the specific form of the
Green function and interpolation used to get the force. The technqiue
is attractive due to its simplicity and the fact that it is
numerically very robust. Highly efficient implementations have been
developed and used during the past decade.  The technique is described
in detail by Hockney \& Eastwood (1981) and we refer the reader to this
book for further details. The specific implementations used in our
study are those of the \ap3m code and the ART code. The PM simulations
presented here have been run using the publicly available \ap3m code
with the particle-particle part switched off and by the ART code with
the mesh refinement block switched off.  In the remainder of this
section we will describe the \ap3m and ART algorithms and the
specifics of our test simulations.

\subsection{AP$^3$M Code}

Particle-Particle-Particle-Mesh (P$^3$M) codes (Hockney et~al. 1973;
Hockney \& Eastwood 1981) express the inter-particle force as a sum of
a short range force (computed by direct particle-particle pair force
summation) and the smoothly varying part (approximated by the
particle-mesh force calculation). One of the major problems for these
codes is the correct splitting of the force into a short-range and a
long-range part. The grid method (PM) is only able to produce reliable
inter particle forces down to a minimum of at least two grid cells.
For smaller separations the force can no longer be represented on the
grid and therefore one must introduce a cut-off radius $r_e$ (larger
than two grid cells !) where for $r < r_e$ the force should smoothly
go to zero.  The parameter $r_e$ defines the chaining-mesh and for 
distances smaller than this cutoff radius $r_e$ a
contribution from the direct particle-particle (PP) summation needs to
be added to the total force acting on each particle. Again this PP
force should smoothly go to zero for very small distances in order to
avoid unphysical particle-particle scattering.  This cutoff of the PP
force determines the overall force resolution of a P$^3$M code.

The most widely used version of this algorithm is currently the adaptive
P$^3$M (\ap3m) code of Couchman (1991). The smoothing of the force in this code
is connected to a $S_2$ sphere, as described in Hockney \& Eastwood (1981).
The particles are treated as density spheres with a profile

\begin{equation} \label{S2}
 S_2: \rho(r) = \displaystyle \left \{
          \begin{array}{ll}
           \displaystyle \frac{48}{\pi \epsilon^4} 
                         \left(
                               \frac{\epsilon}{2} - r
                         \right), 
                                 & \mbox{ for \ } r < \epsilon/2\\
                                 & \\
            \mbox{ \ \ }0,       & \mbox{ otherwise }  \\
          \end{array}
         \right.
\end{equation}

\noindent
where $\epsilon$ is the softening parameter. For distances greater than $\epsilon$ the
particles are treated as point masses interacting according to the newtonian
$1/r^2$ law, whereas for smaller separations the effective shape
of the $S_2$ sphere influences and modifies the force law such 
that the interaction drops down to zero as $r\rightarrow 0$.

When splitting the force into short- and long-range components, one
has to use {\em two} softening parameters: one which is directly
connected to the cut-off radius $r_e$ for the PM force and therefore
tells us where to match the PM and PP part, and another which determines
the overall force resolution (softening scale for PP force). The PP
force is truncated at both the very low separations and at $r \geq
r_e$ where the force can be calculated using the mesh based PM method.
The AP$^3$M code uses a cut-off radius $r_e$ for the long-range force
of approximately 2.4 PM mesh cells, and this leads to the softening
parameter of $\epsilon_{\rm PM} = 1.3 r_e \approx 3.1$ (cf. Hockney
\& Eastwood 1981).  The softening $\epsilon$ of the PP force
determines the overall force resolution; for the \ap3m simulations
presented in this paper the softening scales are given in
Table~\ref{param}. When setting the overall softening parameter to a
value greater than 3.5 the code runs as a pure PM mesh based code,
because the softening of the PP force is greater than that of the PM
part.

Unfortunately, the particle-particle summation which allows one to
achive sub-grid resolutions and thereby makes the P$^3$M algorithm
attractive, is also the method's main drawback. The PP calculation
must search for neighbors out to roughly two mesh spacings to properly
augment the PM force. This becomes increasingly expensive as
clustering develops and particles start to clump together (within the
chaining-mesh cells). The adaptive P$^3$M algorithm remedies this by
covering the high-density, most computationally expensive regions with
refinement grids. Within the refinements the direct sum is replaced by
a further local P$^3$M calculation with isolated boundary conditions
performed on a finer refinement grid (PM mesh and chaining-mesh are
refined). The number of particles per refinement grid cell is smaller
and so is the PP associated computations.  The number of grid cells
per refinement depends on the total number of particles within that
region, but is always a power of two.  The criterion for placing a
refinement depends only on the total number of particles inside a
chaining-mesh cell. If this value exceeds a preselected threshold in a given 
region, the region is refined; for
our runs we used a value of 50 particles. It is convenient to isolate
patches which cover an exact cubic block of chaining-mesh
cells. Recursively placed refinements are allowed, and in the
simulations presented in this paper a maximum level of 3 was reached.

The \ap3m code integrates equations of particle motion using $p=a^{3/2}$ as a
time variable (here $a$ is the expansion factor; Efstathiou et al. 1985).  A
time-centered leapfrog integration with constant time step $\Delta p$
is used.  This scheme leads to ''large'' steps in $a$ at the beginning
of the simulation which then get progressively smaller during the course of
simulation.

\subsection{ART Code}

The Adaptive Refinement Tree code (ART; Kravtsov et al. 1997) reaches
high force resolution by refining all high-density regions with an
automated refinement algorithm.  The refinements are recursive: the
refined regions can also be refined, each subsequent refinement having
half of the previous level's cell size.  This creates an hierarchy of
refinement meshes of different resolutions covering regions of
interest.

The refinement is done cell-by-cell (individual cells can be refined
or de-refined) and meshes are not constrained to have a rectangular
(or any other) shape. This allows the code to refine the required
regions in an efficient manner.  The criterion for refinement is the {\em
local overdensity} of particles: in the simulations presented in this
paper the code refined an individual cell only if the density of
particles (smoothed with the cloud-in-cell scheme; Hockney \& Eastwood
1981) was higher than $n_{th}=5$ particles. Therefore, {\em all}
regions with overdensity higher than $\delta = n_{th}{\
}2^{3L}/\bar{n}$, where $\bar{n}$ is the average number density of
particles in the cube, were refined to the refinement level $L$. For
the two ART simulations presented here, $\bar{n}=1/8$.  The Poisson
equation on the hierarchy of meshes is solved first on the base grid
and then on the subsequent refinement levels. On each refinement level
the code obtains the potential by solving the Dirichlet boundary problem
with boundary conditions provided by the already existing solution at
the previous level. There is no particle-particle summation in the ART
code and the actual force resolution is equal to $\approx 2$ cells of
the finest refinement mesh covering a particular region.  A detailed
description of the code, its tests, and discussion of the force shape
is given in Kravtsov et al.  (1997). Note, however, that the present
version of the code uses multiple time steps on different refinement
levels, as opposed to the constant time stepping in the original
version of the code. The multiple time stepping scheme is described in
some detail in Kravtsov et al. (1998; also see below).

The refinement of the time integration mimics spatial refinement and
the time step for each subsequent refinement level is two times smaller than
the step on the previous level. Note, however, that particles on the
same refinement level move with the same step. When a particle moves
from one level to another, the time step changes and its position and
velocity are interpolated to appropriate time moments. This interpolation
is first-order accurate in time, whereas the rest of the integration is done 
with the second-order accurate time centered leap-frog scheme. All equations
are integrated with the expansion factor $a$ as a time variable and the
global time step hierarchy is thus set by the step $\Delta a_0$ at the
zeroth level (uniform base grid). The step on level $L$ is then
$\Delta a_L=\Delta a_0/2^L$.  

The choice of an appropriate time step for a simulation is dictated by
the peak force resolution. The number of time steps in our simulations
is such that the {\it rms} displacement of particles during a single
time-step is always less than 1/4 of a cell. No particles moves
further than $\sim 0.5$ cells in a single time step, where the cell
size and time step for particles located on the refinement level $L$
are $\Delta x_0/2^L$ and $\Delta a_0/2^L$, respectively.  The value of
$\Delta a_0=0.0015$, used in run ART1 (see Table~\ref{param}) was
determined in a convergence study using a set of $64^3$ particle
simulations described in Kravtsov et al. (1998).  To study the effects
of time step, we have also run a simulation with $\Delta a_0=0.003$.

The ART code integrates the equations of motion in {\em comoving}
coordinates.  Therefore, if a fixed grid is used to calculate the
forces, the force resolution of the simulation degrades as $\propto a
= (1+z)^{-1}$ (see
\S~1).  In order to prevent this and to preserve the initial
resolution in {\em physical} coordinates in the simulations presented
in this paper, the dynamic range between the start ($z_i=87$) and the
end ($z=0$) of the simulation should increase by $(1+z_i)$: i.e., for
our simulations reach $128\times(1+z_i)=11,136$.

In the simulations presented in this paper, the peak resolution is
reached by creating a refinement hierarchy of five levels of
refinement in addition to the base $128^3$ uniform grid. However, the
small number of particles in these simulations does not allow the code
to reach the required target dynamic range of $11,136$, estimated above.

\begin{table}
\caption{Parameters of the numerical simulations. The corresponding
         force resolution in $h^{-1}$kpc is included in parenthesis. Note that PM1 and PM2 
        were simulated using the \ap3m code, while PM3 and PM4 have been simulated 
        using the ART code with mesh refinement switched off. The number of time steps
        for the ART simulations is given for the zeroth level; the effective 
        number of time steps for particles on level $L$ is $\times 2^L$, giving $21120$ 
        steps for the maximum refinement level of the ART1 run.}
\label{param}
 \begin{center}
 \begin{tabular}{|l||ll|c|c|} \hline
{\bf simulation}& \multicolumn{2}{|c|}{softening}
                                               & dyn. range & steps  \\ \hline
 \raisebox{4mm}{}AP$^3$M5 & 0.06 & (7.0) &    2133    & 8000   \\ \hline
 \raisebox{4mm}{}AP$^3$M1 & 0.03 & (3.5) &    4267    & 8000   \\ \hline
 \raisebox{4mm}{}AP$^3$M4 & 0.03 & (3.5) &    4267    & 2000   \\ \hline
 \raisebox{4mm}{}AP$^3$M2 & 0.02 & (2.3) &    6400    & 6000   \\ \hline
 \raisebox{4mm}{}AP$^3$M3 & 0.015& (1.8) &    8544    & 6000   \\ \hline
 \raisebox{4mm}{}ART1  & 0.03125 & (3.7) &    4096    & 660    \\ \hline
 \raisebox{4mm}{}ART2  & 0.03125 & (3.7) &    4096    & 330    \\ \hline
 \raisebox{4mm}{} PM1     & 3.5    & (409) &    128     & 6000   \\ \hline
 \raisebox{4mm}{} PM2     & 3.5    & (409) &    128     & 1500   \\ \hline
 \raisebox{4mm}{} PM3     & 2    & (234) &    128     & 1500   \\ \hline
 \raisebox{4mm}{} PM4     & 2    & (234) &    128     & 750   \\ \hline
 \end{tabular}
 \end{center}
 \end{table}

   \begin{figure*}
   \resizebox{\hsize}{!}{\includegraphics{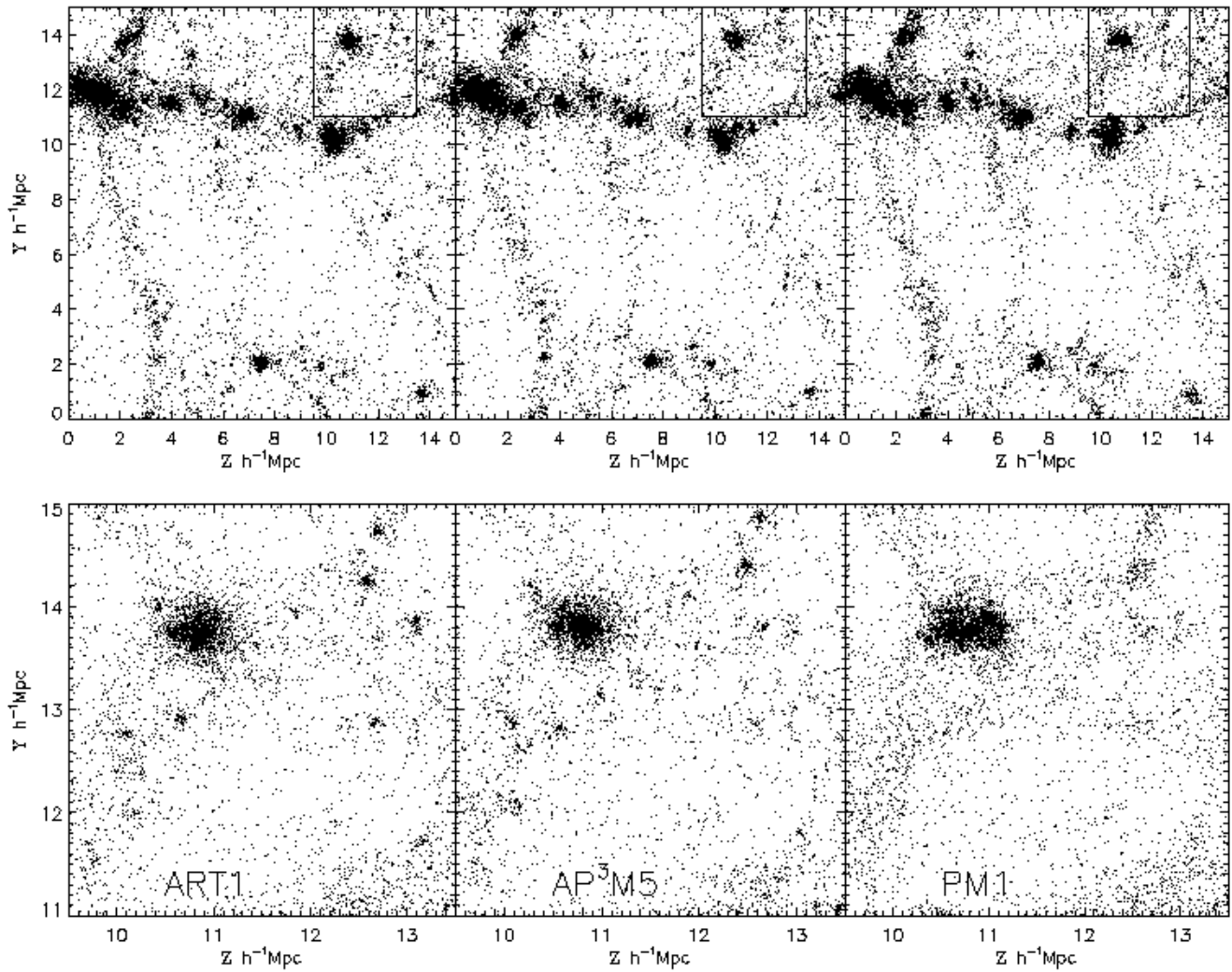}}
   \caption{Slice of 3 $h^{-1}$ Mpc (top row) through the ART (left),
            AP$^3$M (middle), and PM(right) simulation.  In the bottom row we
            zoom into the region marked in the top row.}  
   \label{slice}
   \end{figure*}

\subsection{Simulations}

Although we are not directly interested here in cosmological
applications we decided to use a definite cosmological model: the cluster
normalized, $\sigma_8=0.67$, standard cold dark matter (SCDM) model.
All simulations were run on a $128^3$ mesh with $64^3$ particles and
started with the same random realization at $z=87$. The adopted box
size is $15\hMpc$, which gives a mass resolution of $3.55 \times 10^9
h^{-1} {\rm M}_{\odot}$. The AP$^3$M simulations were carried out
varying both force resolution and time steps.  The two runs of the ART
code differ only in the number of integration steps on the
lowest-resolution of the uniform grid.

For comparison we also ran the PM simulations. Both the \ap3m
and ART have an internal PM block and we ran two pairs of PM
simulations using these different PM implementations. This is done to
compare implementations and to explore the effects of the time
integration scheme on the results.  In the case of \ap3m (PM1 and PM2)
both the adaptive and the PP part of the P$^3$M code have been
switched off, while in the ART PM runs (PM3 and PM4) we have simply
switched off mesh refinement. The parameters of the simulations are
summarized in Table~\ref{param}. The force softening is given in grid
units and in $h^{-1}$kpc (in brackets), and the number of steps of the
ART simulations is presented for the lowest-resolution (the effective
number on the highest-resolution is 32 times larger).

Using this set of runs we can compare simulations with the same
force resolution but different integration steps (e.g., AP$^3$M1 with
AP$^3$M4, ART1 and ART2) amongst each other and simulations with the
same number of integration steps but with varying force resolution
(e.g., AP$^3$M1 with AP$^3$M5). Additionally we compare the different
$N$-body codes in order to quantify the deviations due to different
(grid-based) methods to solve the equations of motion.

It is important to keep in mind that the shape of the small-scale
force is somewhat different in the codes used. Therefore, equal
dynamic range does not correspond to the same physical resolution. The
peak resolution of the ART code is $\approx 2$ cells of the highest
level refinement, and so the actual force resolution is twice worse
than the ``formal'' resolution given by the dynamic range.  To make
cross-code comparison, we have performed the simulation AP$^3$M5,
which has approximately half the dynamic range of the ART runs and
similar force resolution.

\section{Properties of the particle distribution}

   \begin{figure} \resizebox{\hsize}{!}{\includegraphics{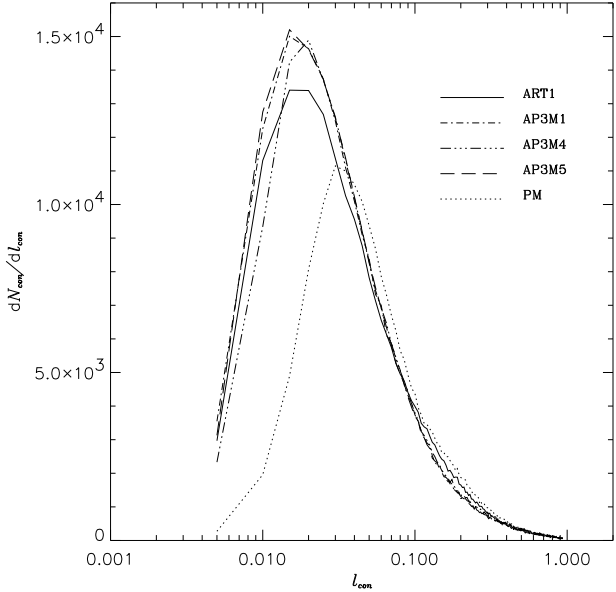}}
   \caption{Number of connections of the minimum spanning tree per bin
   of the connection lengths $l_{con}$ as a function of the connection length.
   The connection length is in units of the mean interparticle distance $\bar
   l$, the bin size is $dl_{con}=0.005*\bar l$.}  \label{mst} \end{figure}

   \begin{figure}
   \resizebox{\hsize}{!}{\includegraphics{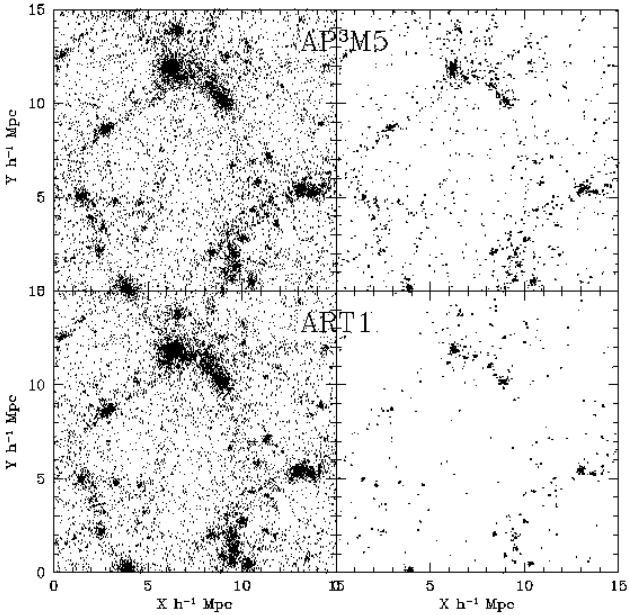}}
   \caption{Left: Projections of 5\% of all particles of the box along
   the z-axis. Right: Projections of all doublets and triplets found
   with $ll = 0.015$. Note that the systems linked using this linking
   length have overdensity of $\delta\grtsim 420000$.}  \label{doublets}
   \end{figure}

\subsection{Visual comparison}

A first inspection of the global distribution of particles in a 
3$h^{-1}$Mpc thick slice (Fig.~\ref{slice}, top row) shows that the
distributions are very similar. Even the much lower resolution PM1
simulation shows virtually the same global particle distribution. In
the bottom row of the figure we zoom into the region marked by
rectangles in the upper panel. Here one can clearly see that the two
high resolution runs produce many small-size dense halos, which are, however, 
slightly shifted in their positions. We attribute this shift to the 
cumulative phase errors due to the differences in the time integration 
schemes of the type observed and discussed in a recent code comparison 
by Frenk et al. (1999; Santa-Barbara cluster comparison project). 
Most of these small clumps are absent in the PM run due to its poor 
resolution. The small scale density peaks do not collapse on scales 
smaller than the 2 PM grid cells due to the sub-newtonian of self-gravitation
at these scales (see \S~3.3). From Figure~\ref{slice} it can be seen that only halos
of size larger than approximately one grid cell collapse in the PM simulation
(see, e.g., Klypin 1996). 

Our comparison can be contrasted with comparison by Suisalu
\& Saar~(1996) (Fig.1 and Fig.2 in their paper). If compared 
on a large scale, the particle distributions in different runs compare
much better in our case than the simulations compared in Suisalu \&
Saar (1996). This indicates that the differences they observed are due
to the differences in their multigrid algorithm rather than to the
two-body scattering, as argued in their paper.

\subsection{The minimal spanning tree}

To quantify the differences between the simulations we have calculated
the minimal spanning tree (MST) of the particle distribution. The
minimal spanning tree of any point distribution is a {\em unique}, well
defined quantity which describes the clustering properties of the
point process completely (e.g., Bhavsar \& Splinter 1996, and
references therein). The minimal spanning tree of $n$ points contains
$N-1$ connections. For the ART1, AP$^3$M1, AP$^3$M4, AP$^3$M5 and a PM1
simulations we show in Fig.~\ref{mst} the number of connections
$N_{con}$ of the tree per bin of the connection length (equal to
$0.005*\bar l$) as a function of the length of the connection,
$l_{con}$. Here, $\bar l = (V_{box}/N)^{1/3}$ denotes the
mean inter-particle separation. Since the length of the connection is
proportional to $\rho^{-1/3}$ the probability distribution of
connections ($N_{con}/N$) is equivalent to the density probability
distribution in the simulation. 

In Fig.~\ref{mst}, the connection length distributions for ART and
AP$^3$M simulations are peaked at the same relative connection length
($\approx 0.015-0.02 \bar{l}$), whereas the PM simulation is peaked at
a higher value ($\approx 0.03-0.04 \bar{l}$). This indicates the
ability of the ART and the AP$^3$M codes to resolve shorter scales
and, therefore, to reach higher overdensities.  The position of the
maximum depends only slightly on the resolution. In fact, increasing
the resolution from PM to ART by a factor of 32 shifts the maximum by
about a factor of 2. This is probably because the differences affect
only a very small fraction of the volume and dark matter particles.
Correspondingly, differences in resolutions of AP$^3$M1 and AP$^3$M5
are too subtle to have a visible effect on the
distribution. 

A time integration has a much more noticeable effect.  An increase of the
integration step in the run AP$^3$M4 compared to AP$^3$M1 and AP$^3$M5
leads to a shift of the maximum (and distribution) to higher values of
$l_{con}$. This is caused by inaccuracies in the integration of particle
trajectories in high-density regions, as is evidenced, for example, by
the halo density profiles in the AP$^3$M4 run (see also results below).

The maxima of the AP$^3$M simulations are slightly higher in amplitude
than the maximum of the ART simulation. This reflects the fact that
the AP3M resolves forces uniformly (i.e., equally well in both low- and
high-density regions).  The ART code by design reaches high resolution
only in the high-density regions. Therefore, there are groups of a few
particles located in low-density environments and thus not resolved by
the ART code, which are however resolved by the \ap3m. For example, we
have found 3618 doublets and 558 triplets linked by $ll = 0.015$ (the
maximum of the distribution) in the AP$^3$M5 run, whereas only 2753
doublets and 466 triplets are found in the ART1 simulation at this
linking lengths.

Figure~\ref{doublets} demonstrates that many of the missing doublets
and triplets are indeed located in the low-density regions. The right
column of this figure shows the projection of the distribution of all
doublets and triplets found in the AP$^3$M5 (top) and ART1 (bottom)
runs by the friends-of-friends algorithm with the linking length of
$ll = 0.015$. For comparison, we show in the left column of
Fig.~\ref{doublets} a projections of randomly selected 5\% of all
particles.  While some doublets in the low-density regions are found
in the ART run, all of them are gravitationally unbound (are chance
superpositions). We find that in the \ap3m runs most of such doublets
and triplets are bound and are a part of a binary, triple, or higher
multiplicity clusters.

According to the sampling theorem, one needs at least $\sim 20-30$
particles to resolve three-dimensional waves in the initial power
spectrum. The presence of gravitationally bound clusters consisting of
just a few particles is therefore artificial.

\begin{table}
\caption{Density cross-correlation coefficient 
         $K = \displaystyle \frac{\langle\delta_1 \delta_2\rangle}{\sigma_1 \sigma_2}$.}
\label{denscorr}
 \begin{center}
 \begin{tabular}{|lcl||c|c|c|c|c|} \hline
      &                     &           & \multicolumn{5}{c|}{\bf grid} \\ \hline
  \multicolumn{3}{|c||}{\bf simulation}&   32  & 64   & 128  & 256  & 400  \\ \hline \hline
  \raisebox{4mm}{}ART1     & $\leftrightarrow$ & ART2
                 & 0.98  & 0.98 & 0.94 & 0.84 & 0.72 \\ \hline
  \raisebox{4mm}{}AP$^3$M1 & $\leftrightarrow$ & AP$^3$M4     
                 & 0.98  & 0.98 & 0.98 & 0.93 & 0.87 \\ \hline
  \raisebox{4mm}{}PM1      & $\leftrightarrow$ & PM2    
                 & 1.00  & 1.00 & 1.00 & 1.00 & 1.00 \\ \hline   
  \raisebox{4mm}{}PM3      & $\leftrightarrow$ & PM4    
                 & 1.00  & 1.00 & 1.00 & 1.00 & 1.00 \\ \hline \hline

  \raisebox{4mm}{}AP$^3$M2 & $\leftrightarrow$ & AP$^3$M3 
                 &  0.98 & 0.98 & 0.98 & 0.93 & 0.87 \\ \hline
  \raisebox{4mm}{}AP$^3$M1 & $\leftrightarrow$ & AP$^3$M5 
                 &  0.98 & 0.99 & 0.99 & 0.96 & 0.91 \\ \hline   
\hline

  \raisebox{4mm}{}PM2      & $\leftrightarrow$ & PM3    
                 & 1.00  & 0.98 & 0.93 & 0.84 & 0.73 \\ \hline 
  \raisebox{4mm}{}ART1     & $\leftrightarrow$ & AP$^3$M1      
                 & 0.95 & 0.88 & 0.71 & 0.44 & 0.30 \\ \hline
  \raisebox{4mm}{}ART1     & $\leftrightarrow$ & AP$^3$M5      
                 & 0.95 & 0.88 & 0.71 & 0.44 & 0.30 \\ \hline
  \raisebox{4mm}{}ART1     & $\leftrightarrow$ & PM1
                 & 0.96 & 0.92 & 0.80 & 0.60 & 0.47 \\ \hline
  \raisebox{4mm}{}AP$^3$M1 & $\leftrightarrow$ & PM1
                 &  0.95 & 0.90 & 0.79 & 0.61 & 0.48 \\ \hline
 \end{tabular}
 \end{center}
\end{table}

\subsection{Density Cross--Correlation Coefficient}

The density cross--correlation coefficient,
\begin{equation}
 K = \displaystyle \frac{\langle\delta_1 \delta_2\rangle}{\sigma_1 \sigma_2},
\end{equation}
was introduced by Coles et al. (1993) in order to quantify similarities
and differences between simulations of different cosmological models.
Recently, Splinter et~al. (1998) have adopted this statistic to
compare simulations. Here, we follow the same approach and use this
measure to quantify differences between simulations which have been
carried out by different numerical algorithms or by the same algorithm
but with different parameters of the simulation.

To compute $K$, we have calculated the densities on a regular mesh
using the triangular-shaped cloud (TSC; Hockney \& Eastwood 1981)
density assignment scheme and then used the resulting density field to
compute $\langle\delta_1 \delta_2\rangle$ and variances. We have varied the size
of the grid in order to show the dependence of the cross-correlation
on the smoothing scale of the density field.  

We summarize our results in Table~\ref{denscorr}: the first four rows
present the cross-correlation coefficients between the runs with the
same time integration scheme but different time steps. In the
following two rows we present the cross-correlation coefficients for
the \ap3m runs with different force resolutions but the same
integration step. The rest of the rows in the table present the
cross-correlation coefficients for the runs with different time
integration schemes as well as the cross-correlation coefficients
between the AP$^3$M1 and PM1 runs which have been simulated with the
same time itegration scheme but with vastly different force
resolutions. 

In all cases, it is obvious that the cross-correlation worsens for
smaller smoothing scales (the larger density grid size).  With smaller
smoothing, smaller structures in the density field are resolved. The
degraded cross-correlation therefore indicates that there are
differences in locations and/or densities of these small-scale
structures. It is also clear from the definition of $K$ that this
measure is particularly sensitive to the differences in the highest
density regions.  If we restrict the correlation analysis to a coarse
grid, we smooth the particle distribution with a fairly large smoothing
length and smear out the details and differences of the small-scale
structure.

The differences revealed by the cross-correlation coefficient can
arise either because the internal density distribution of the
structures is different in different runs or because the spatial
locations of these structures are somewhat different. Our analysis of
halo profiles (See
\S~\profiles) shows that differences in density in the same halos are
small (except, of course for the PM runs) and have no significant
effect on the cross-correlation. The degrading cross-correlation in
high-resolution runs is thus due to the differences in the locations
of small-scale structure rather than to the differences in
density. Indeed, one can readily see in the bottom row of
Fig.~\ref{slice} that positions of small clumps in the ART and \ap3m
simulations often differ by $\sim 100-300\hkpc$, the scale at which
a significant decrease in $K$ is observed. With such shifts in halo
locations, the same halo may occupy different cells in the density
grid which systematically reduces $\langle\delta_1
\delta_2\rangle$ (calculated cell-by-cell) and results in a lower $K$.

What causes the differences in halo positions? The rows 1, 2, 5, and 6
of Table~\ref{denscorr} indicate that both force resolution and time
step have the same effect on $K$, both causing some phase errors.
However, rows 7-9 show that the time integration scheme causes much
bigger phase differences than either force resolution or time step.
This was indeed observed in the recent ``Santa Barbara Cluster''
code comparison project (Frenk et al. 1999). Different time integration
schemes lead to a different accumulation of the phase errors. The manifestation
of these differences is certain ``asynchronicity'' between simulations:
the same phase error is accumulated at slightly different time moments. This 
results in shifts of halo positions when simulations are compared at the same 
time moment. 

We have indeed observed such asynchronicity in our simulations. Thus,
for example, cross-correlation coefficient $K$ between ART1 and
AP$^3$M5 on a $128^3$ grid reaches a maximum of 0.84 at $a=1.04$ when
AP$^3$M5 is evolved further in time (this can be compared to 0.71 at
$a=1.0$ in Table~\ref{denscorr}) while keeping ART1 fixed at a=1.0.
Similar effect is observed at $z=2$: the maximum $K$ is achieved when
AP$^3$M5 is advanced forward in time by a factor of 1.02 in expansion
factor. Partly, this asynchronicity may be caused by the initial phase
error introduced at the start as the {\ap3m} simulations particles
were advanced half a step (or by a factor of 1.03 in expansion factor)
forward. The additional error accumulates during the time integration.

While most halo properties and properties of the matter
distribution appear to be similar in the ART and the \ap3m runs (see
results below), the differences between positions of small-scale
structures are much larger between these runs than any differences
between runs simulated using the same code. This is clearly seen in
the case of PM runs. All 4 runs cross-correlate perfectly within the
code type (PM1 and PM2 were run using \ap3m, while PM3 and PM4 were
run using the ART code), but cross-correlate rather poorly when
different code simulations are compared. In the latter case we
observed a decrease of $K$ as we go to smaller smoothing scales
similar to that observed in other cross-code cross-correlation
coefficients.

One should of course bear in mind that the
shape and accuracy of the PM force in {\ap3m} and ART codes are
somewhat different at small ($\sim 1-4$ grid cells) scales. In the
{\ap3m}, the PM force is shaped using modification of the Green
functions (Couchman 1991) which is controlled by a special softening
parameter $\epsilon_{\rm PM}$.  This procedure considerably increases accuracy
of the force (down to $\lesssim 5\%$ in the case of $\epsilon_{\rm PM}\approx
3.5$ used in our PM1 and PM2 simulations) in the force, at the expense
of making the force somewhat ``softer''. Thus, for example, in PM1 and
PM2 runs, the force becomes systematically smaller than the Newtonian
value at separations $\lesssim 3$ grid cells (15\% smaller at $r=2$
cells and 70\% smaller at $r=1$ cell). The Green functions in the PM
solver in the ART code are not modified to reduce the errors. This results in
the force which is Newtonian on average down to the scale of one grid
cell (the force then falls off sharply at smaller separations).  At
the same time, the errors at small separations are considerably higher
(see Gelb 1992; Kravtsov et al. 1997). At separation of 1 grid cell,
the errors may reach $\sim 50\%$ (albeit in small number of particle
configurations). About $10-20\%$ of that error is due to the cubical
shape of the particles assumed in the PM algorithm, while the
remaining error arises from numerical differentiation of the
potential.  These errors, however, are not systematically positive or
negative but are scattered more or less evenly around zero. This means
that particle trajectories can be integrated stably down to
separations of 1-2 grid cells, as was demonstrated in Kravtsov et
al. (1997).

Although it is important to keep the differences in force shape and accuracy
in mind, we think that they are not the main cause of poor cross-correlations.
The halo density profiles in different PM runs are in good agreement at all 
resolved scales and thus differences in internal density distributions 
cannot explain low cross-code $K$. At the same time, visual comparisons 
of halo positions show small shifts that are most naturally explained by 
the different phase errors accumulated in different codes. 

This calls into question the usefulness of cross-correlation
coefficient for studies of resolution effects (Splinter et al. 1998),
unless the study is done within the same numerical code.

This conclusion can, in fact, be drawn from results of Splinter et
al. (1998): when the simulation evolves into the highly nonlinear stage the
cross-correlation within the same code is much better than
cross-correlation between runs of similar resolution but simulated
using different codes. For example their Table 4 shows that on a
$128^3$ grid coefficient $K$ for the TREE-code runs with the force
resolutions of $\epsilon =0.0625$ and $\epsilon =0.25$ (in the units
of mean interparticle separation) is 0.87, while the cross-correlation
coefficient between the TREE $\epsilon =0.0625$ and \ap3m $\epsilon
=0.25$ runs is only $0.67$. This is similar to the value of $0.71$
which we obtained for $K$ on the same grid size for ART1 and \ap3m
runs. Moreover, Figure~9 in Splinter et al. (1998) also agrees with 
our conclusion. The phase errors demonstrated in this figure increase 
towards smaller scales which explains the decrease of cross-correlation
coefficient for finer grids. Moreover, the figure also shows that 
cross-correlation {\em within single code type} is always good regardless
of the mass or force resolution. The largest phase differences are 
observed between runs simulated using different codes. 

The differences in time integration schemes are not the only possible
sources of small-scale differences in density fields. For example, the
last two rows of Table~\ref{denscorr} show that cross-correlation
between high-resolution ART and \ap3m runs and low-resolution PM run
is poor regardless of whether the integration scheme was the same
(AP$^3$M1 $\leftrightarrow$ PM1) or different (ART1 $\leftrightarrow$
PM1). In this case, the differences in the small-scale details of density
fields are due to the vastly different force resolutions of the
simulations.  As was noted above, the low resolution of the PM
simulation ($234\hkpc$) precludes collapse of any halos with the size
smaller than $\sim 1-2$ grid cells (see Fig.~\ref{slice}). In the locations
of small-size halos $\delta$ is very high for ART and \ap3m runs but
is much lower (because there are no halos) in the PM run; hence the
considerably lower cross-correlation coefficient.

Given the above considerations, our interpretation of our results and
the results of Splinter et al. (1998) is markedly different than
the interpretation by the authors of the latter study. These authors
interpret the differences between the low- and high-resolution runs as
an erroneous evolution in the latter, whereas our interpretation
(obvious from Fig.~\ref{slice}) is that these differences are due to the
fact that high-density small-scale structures such as halos do not
collapse in low-resolution runs. The differences between
high-resolution runs are interpreted as the phase errors leading to
small shifts to the locations of small-scale structures, as discussed
above and as observed in other studies (Frenk et al. 1999).

The origin of these phase errors is the dynamical instability of particle
trajectories in the high-density regions. As is well known, the
trajectories in the virialized systems tend to be chaotic and any
small differences existing at any time moment will tend to grow very
fast with time. The divergence can thus be expected to be more
important in nonlinear regions and this explains the larger phase
errors at smaller scales. The differences may be caused by the
difference in the force calculation, errors introduced by numerical time
integration, or simply by different roundoff errors. The resulting
phase errors are cumulative and thus will grow with time.

Unfortunately, it is impossible to tell which code has the smallest
phase errors, because we do not know how the phases should evolve in
the high-density regions. However, this is probably not the point. 
Phase errors of this kind will be very difficult to get rid of,
because even in the case of infinitely good mass, force, and time
resolutions, there will always be roundoff errors, which will behave
differently in different codes and thus will tend to grow differences
in phases. Luckily, almost all popular statistics used in cosmological
analyses are not sensitive to phases and therefore results are not
affected by this problem. Moreover, it is clear that at scales
$\grtsim 1\hMpc$ the errors in phases become negligible (different
runs cross-correlate perfectly) and therefore even phase-sensitive
analyses should not be affected if restricted to large
scales. However, it is clear that the existence of such errors should be
kept in mind when analyzing or comparing cosmological simulations.

   \begin{figure}
   \resizebox{\hsize}{!}{\includegraphics{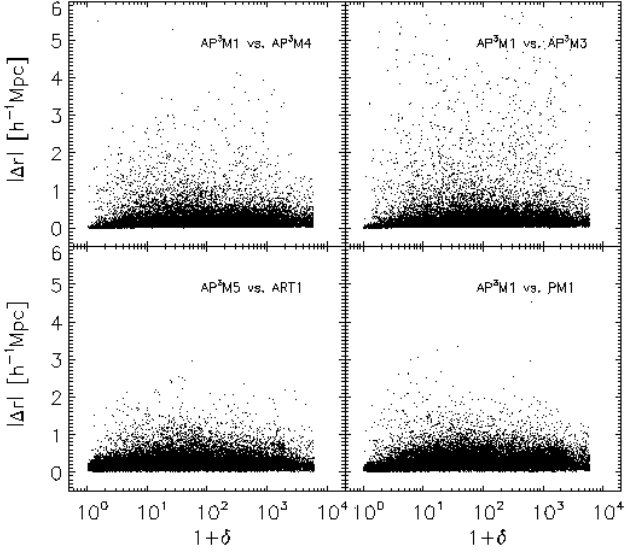}}
   \caption{Deviation of particle coordinates at $z=0$ in two
   simulations (started with identical initial particle distribution) depending
   on the local overdensity (estimated using $128^3$-cell grid) at the
   location of particle in the first simulation. Only 10\% of all
   particles are shown. In the upper panel of Fig.~\ref{drdens1} we
   compare runs simulated using the same code but different
   integration steps (left) and different force resolution (right). In
   the lower panel we compare runs simulated using different
   codes.}  \label{drdens1} \end{figure}

   \begin{figure}
      \resizebox{\hsize}{!}{\includegraphics{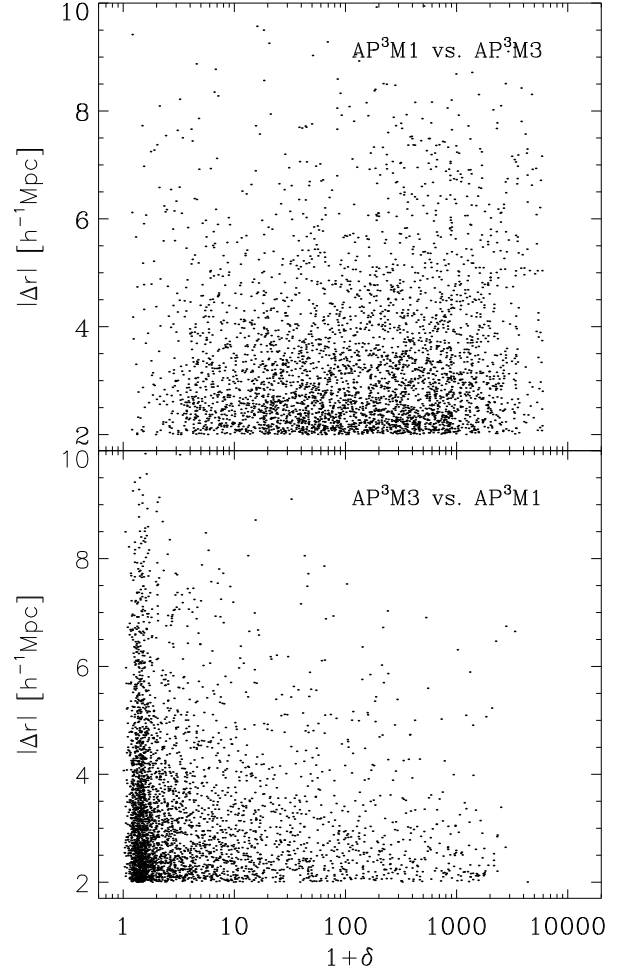}}
      \caption{The same as in Fig.~\ref{drdens1} but for the runs
      AP$^3$M1 and AP$^3$M3. In this case {\em all\/} particles with
      $\vert \Delta r\vert>2\hMpc$ are shown (these particles
      constitute $\approx 1.4\%$ of the total number of particles).
      The overdensity in the upper panel is estimated at the location
      of the particle in the AP$^3$M1 run, while in the bottom panel it is
      estimated for the corresponding particle in the AP$^3$M3
      run. Note that counterpart particles in AP$^3$M3 tend to be
      located in low-density regions.}  \label{drdens2} \end{figure}

   \begin{figure*}
      \resizebox{\hsize}{!}{\includegraphics{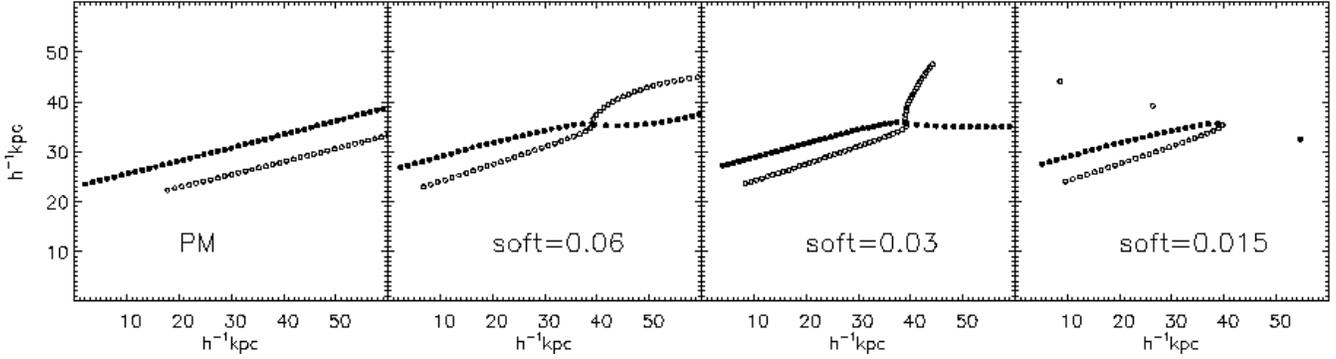}}
      \caption{Trajectory of two particles in
               AP$^3$M simulation for different force
               resolutions ($\Delta t$=const). The figure demonstrates 
               the presence of two-body scattering in all of our \ap3m 
               simulations.}
      \label{ipartlocal}
    \end{figure*}

\subsection{Particle trajectories}

As we have discussed in the previous section, small numerical errors
tend to grow and lead to deviations of the particle
trajectories in nonlinear regions. Nevertheless, it is clear that the
maximum deviations should be approximately equal to the size of a typical
halo. Although particle trajectories can deviate, they are expected to stay
bound to the parent halo. There is an additional deviation of the
order of $\sim 100-300\hkpc$ in the positions of the halos themselves
(due to the phase errors), but this is also of the order of halo size.
All in all, we can expect a scatter in the positions of the same 
particles in different simulations not larger than the size of the 
largest systems formed: $\sim 1-2\hMpc$. 

To compare the particle trajectories in our simulations, we have
calculated the deviations of the coordinates 
\mbox{$\Delta r = |\vec{r}^{(1)(k)} - \vec{r}^{(2)(k)}|$}, 
where $\vec{r}^{(i)(k)}$ is the position of the k-th particle in the
i-th simulation.

In Fig.~\ref{drdens1} we have plotted $\Delta r $ as a function of the
local overdensity at the position of the particle
$\vec{r}^{(1)}$ for 10\% of particles randomly selected from the total
number of particles. In the upper panel of Fig.~\ref{drdens1} we compare
runs simulated using the same code but different integration steps (left)
and different force resolution (right). In the lower panels we compare
runs simulated using different codes.

A quick look at Fig.~\ref{drdens1} shows that the scatter in particle
positions is substantial but in most runs it is contained within
$\lesssim 2\hMpc$, the approximate size of the largest system seen in
Fig.~\ref{slice}. The scatter for {\em most} particles in cross-code
differences shown in the bottom row is somewhat larger due to the
larger differences in halo positions. As mentioned above, this
difference adds to the difference in particle positions {\em within\/}
halos.  However, the scatter is even larger when AP$^3$M1
and AP$^3$M3 are compared. In this case, about 1.4\% of particles 
are separated by more than $2\hMpc$. We did not find such outliers
when comparing ART runs or AP$^3$M runs with lower resolution amongst
themselves.  The comparison of AP$^3$M5 and ART1 shows that the
scatter is much smaller.

Figure~\ref{drdens2} shows {\em all\/} the particles with 
separations $\vert \Delta r\vert > 2\hMpc$ in the AP$^3$M1 and
AP$^3M$3 runs. The overdensity in the upper panel is estimated at the
location of the particle in the AP$^3$M1 run, while in the bottom panel it
is estimated for the corresponding particle in the AP$^3$M3 run. It is 
clear that counterpart particles in the AP$^3$M3 tend to be located in
low-density regions, whereas the corresponding particles in the AP$^3$M1
run are located in a wide range of environments.

   \begin{figure}
   \resizebox{\hsize}{!}{\includegraphics{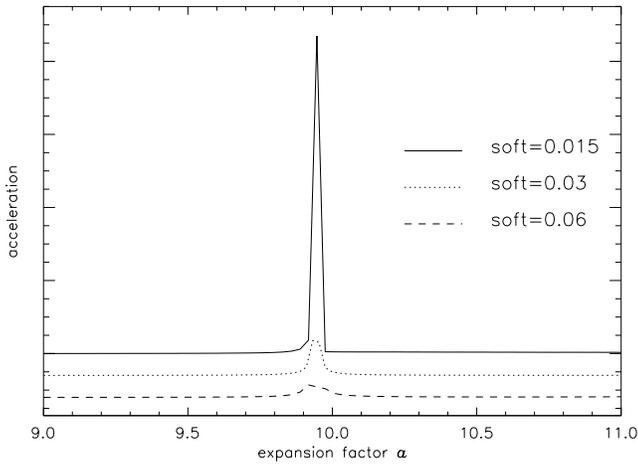}}
   \caption{Acceleration (in arbitrary units) of particle denoted with
   filled circles in Fig.~\ref{ipartlocal}. Dotted and dashed curves
   lowered by a constant offset.}  
   \label{ipartaccel} \end{figure}

The fact that these large deviations occur preferentially in the
highest resolution runs, immediately raises suspicion that they are due
to two-body scattering. Indeed, when we analyzed the trajectories
of the deviant pairs, we found obvious scattering events.  In
Fig.~\ref{ipartlocal} we present an example of such two-body
scattering. In this figure the force resolution increases (softening
decreases) from right to left. For clarity only every second
integration step is shown. 

The region shown in Fig.~\ref{ipartlocal} is smaller than the grid
cell size, so there is no interaction at all in the PM
simulation. The two particles move in the mean potential of the other
particles.  The same happens in the ART simulations. On the contrary,
due to the high force resolution these two particles interact and
approach in the case of \ap3m 1-4 runs. In the case of the highest
force resolution this leads to an interaction which, due to the
insufficiently small time step, leads to a violation of energy
conservation. The particles approach very closely, feel a large force,
undergo in this moment a huge acceleration and move away in opposite
directions with high velocities.  At the next integration step the
particles are already too far to feel substantial two-body
interaction. Therefore, they move with almost constant high velocity
in opposite directions. The velocity is about a factor of 10 higher
than the initial velocity, i.e. the total kinetic energy of the system
increased during the interaction by a factor of 100.  Such pairs of
particles are located at large $\Delta r$ in scatter plots shown in
Figs.~\ref{drdens1}~\&~\ref{drdens2}. We have run an additional
simulation with the parameters identical to those of the AP$^3$M3 run
but with a much larger number of time steps (48,000 steps in total). The
outlying particles dissappear and the plots look similar to the
AP$^3$M5-ART1 plot. This means that although the scattering is still
present, there is no violation of energy conservation in the smaller
step run and therefore there are no high-velocity streaming particles.

Fig~\ref{drdens2} indicates that scattered particles attain large
velocities and move out of high-density regions. Most of the
simulation volume is low-density, so it is not surprising that we find
these streaming particles preferentially in the low-density
regions. Their counterparts in the AP$^3$M1 runs, on the other hand,
are located in a wide range of environments.  This implies either that
the scattered particles were ejected from high-density regions, or
that these particles attained their excess energy in two-body
encounters in low-density regions and simply did not participate in
the gravitational collapse due to their high velocities. 

Fig.~\ref{ipartaccel} shows the acceleration of the particle denoted
by filled circles in the three AP$^3$M plots in
Fig.~\ref{ipartlocal}. This figure illustrates the spike in the
particle acceleration during the scattering event. It also shows that
the higher the force resolution, the larger the acceleration.

Such collisions are possible if the force resolution is independent of
the local particle density. This problem does not arise in the ART
code because the resolution is only increased in the regions of high
local particle density.  We have checked all particles beyond
separation $\Delta r
\mathrel{\hbox{\rlap{\hbox{\lower4pt\hbox{$\sim$}}}\hbox{$>$}}} 2
h^{-1}$Mpc within the both ART runs and could not find any event
comparable to the collision in the AP$^3$M runs.

The conditions for two-body scattering can be estimated by noting 
that strong scattering occurs when the potential energy of two-body 
interaction is equal to kinetic energy of the interacting particles. 
This gives the scale $s=2Gm/v^2$ or 
\begin{equation}
s = 8.61\times 10^{-2}h^{-1}{\ \rm kpc}\left(\frac{m_p}{10^8h^{-1}
{\ \rm M_{\odot}}}\right)
\left(\frac{v}{100{\ {\rm km/s}}}\right)^{-2}.
\end{equation}
For the simulations presented here 
($m_p=3.55\times 10^9h^{-1} {\ \rm M_{\odot}}$)
this scale is
\begin{equation}
s = 3.05\times v_{100}^{-2}{\ }h^{-1}{\ \rm kpc},
\end{equation}
where $v_{100}$ is velocity in units of $100{\ \rm km/s}$. 
In terms of the mean interparticle separation, $d=\bar{n}^{-1/3}$, this gives
\begin{equation}
\tilde{s}\equiv s/d = 0.013 \times v_{100}^{-2}.
\end{equation}

Two-body scattering occurs if the force resolution, $\epsilon$, 
is less than the scale $s$ {\em and} if $s$ is much smaller than the 
{\em local} interparticle separation $d_{loc}$. The latter for these 
simulations is $d_{loc}=234.38(1+\delta)^{-1/3}\hkpc$. 

The above equations show that scattering is possible in the AP$^3$M runs  1 
through 4. This in agreement with results presented in 
Figures~\ref{drdens1} and~\ref{ipartaccel}.

\subsection{2-point correlation function}

\begin{figure}
      \resizebox{\hsize}{!}{\includegraphics{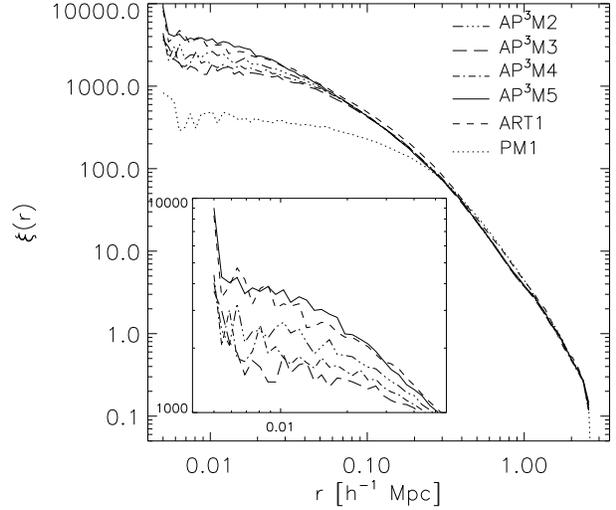}}
      \caption{Correlation function of dark matter particles. Note that the range 
                of correlation amplitudes is different in the inset panel.}
      \label{DMxi}
   \end{figure}

In Fig.~\ref{DMxi} we show the correlation function for the dark
matter distribution down to the scale of $5\hkpc$, which is close to
the force resolution of all our high-resolution simulations. The
correlation function in runs AP$^3$M1 and ART2 are similar to those of
AP$^3$M5 and ART1 respectively and are not shown for clarity. We can
see that the AP$^3$M5 and the ART1 runs agree to $\lesssim 10\%$ over
the whole range of scales. The correlation amplitudes of runs
\ap3m~2-4, however, are systematically lower at $r\lesssim 50-60\hkpc$
(i.e., the scale corresponding to $\approx 15-20$ resolutions), with the
AP$^3$M3 run exhibiting the lowest amplitude. At scales $\lesssim
30\hkpc$ the deviations from the ART1 and the AP$^3$M5 runs are
$\approx 100-200\%$. We attribute these deviations to the 
numerical effects discussed in \S~5. The fact that the AP$^3$M2
correlation amplitude deviates less than that of the AP$^3$M3 run,
indicates that the effect is very sensitive to the force resolution.

The correlation function of the PM simulation deviates strongly on
small scales. However, the bend coincides with the force resolution
($\approx 234\hkpc$) of this run. At scales smaller than resolution, we
can expect an incorrect correlation amplitude because waves of wavelength
smaller than the resolution do not grow at the correct rate in these
runs.

This result agrees with the correlation function comparison done by
Col\'{\i}n et al. (1999), where agreement of $\lesssim 10\%$ was found
between the correlation functions from the larger $256^3$-particle ART,
\ap3m, and PM simulations at all resolved scales. There is no evidence
therefore, that high-resolution simulations, given that they are done
with sufficiently small time step, simulate the 2-point correlation
function incorrectly at scales smaller than the mean interparticle
separation. The agreement between high-resolution and PM simulations
of the same mass resolution always agree at the scales resolved in the
PM runs.

\section{Properties of Halos}

\subsection{Identification of Halos}

To compare properties of the halos and their distribution in different
simulations, we identify DM halos using the the friends-of-friends
(FOF) algorithm. The algorithm identifies clumps in which all
particles have neighbors with distances smaller than $ll$ times the
mean inter particle separation, $r_{ll} \leq ll \times \bar l$.  This
halo finding algorithm does not assume a special geometry for the
identified objects. A drawback is that it only uses particle positions
and therefore can identidy spurious unbound clumps at low halo masses.

The mean overdensity of a particle cluster is related to the linking
length used to identify it. In an Einstein-de-Sitter universe
(simulated in our runs) virialized halos have an overdensity of
$\delta_{{\rm vir}} \approx 178$, which corresponds approximately to
the linking length of $ll=0.2$. A reduction of the linking length by a
factor of 2 roughly corresponds to an increase of the overdensity by a
factor of 8. Using smaller linking lengths we can study the
substructure of the DM halos. Table~\ref{xiparam} lists the number of
halos identified using different linking lengths in different runs for
the two lower limits on the number of particles in a halo: 25 (columns
$2-4$) and 50 (columns $5-7$).

\begin{table}
\caption{Number of DM halos identified using the FOF algorithm as a function
         of the linking length (in units of mean interparticle
         separation). Only halos containing more than $25$ (columns
         $2-4$) and $50$ (columns $5-7$) particles have been counted.}
\label{xiparam}
 \begin{center}
 \begin{tabular}{|l||c|c|c||c|c|c|} \hline
                          &\multicolumn{3}{|c|}{$N_p>25$} 
                          &\multicolumn{3}{|c|}{$N_p>50$}     \\ \hline
 linking length:          &  0.2  & 0.1  & 0.05 & 0.2 & 0.1 & 0.05  \\ \hline \hline
 \raisebox{4mm}{}AP$^3$M5 &  388  & 292  & 162  & 206  & 152  &  85 \\ \hline
 \raisebox{4mm}{}AP$^3$M1 &  377  & 264  & 160  & 204  & 149  &  84 \\ \hline
 \raisebox{4mm}{}AP$^3$M4 &  340  & 221  & 110  & 187  & 120  &  65 \\ \hline
 \raisebox{4mm}{}AP$^3$M2 &  365  & 251  & 122  & 188  & 132  &  74 \\ \hline
 \raisebox{4mm}{}AP$^3$M3 &  332  & 215  &  87  & 173  & 117  &  55 \\ \hline
 \raisebox{4mm}{}PM1      &  214  & 150  &  50  & 134  &  74  &  19 \\ \hline 
 \raisebox{4mm}{}PM2      &  208  & 155  &  49  & 129  &  66  &  18 \\ \hline 
 \raisebox{4mm}{}ART1     &  359  & 291  & 161  & 199  & 148  &  89 \\ \hline
 \raisebox{4mm}{}ART2     &  383  & 297  & 169  & 206  & 156  &  87 \\ \hline
 \end{tabular}
 \end{center}
\end{table}

The table shows that there are differences of $\sim 15-50\%$ between
high-resolution runs. The differences are present not only at low
linking lengths, but even at the ``virial'' linking length, $ll=0.2$.
These differences are partly due to the nature of the FOF algorithm:
small differences in the particle configurations may lead to an
identification of a single halo in one simulation and to the
identification of two or more halos in another
simulation. Nevertheless, some systematic differences are also
apparent. PM simulations have almost half as many halos due to the
absence of small-mass halos (see Fig.~\ref{slice}) that do not
collapse (or do not survive) due to the poor force resolution of these
runs.

Also, while AP$^3$M1, AP$^3$M5, ART1, and ART2 runs agree reasonably
among themselves, the number of halos in runs AP$^3$M2, AP$^3$M3, and
AP$^3$M4 is systematically lower. In this case the differences seem to
be counter-intuitive: the number of halos is smaller in higher force
resolution runs. These differences persist at all linking lengths
indicating that there are differences in substructure as well as in
the number of isolated halos. Moreover, the differences persist
even for a larger cut in the number of particles.  It has been noted in
previous studies (e.g., van Kampen 1995; Moore, Katz \& Lake 1996)
that particle evaporation due to two-body scattering (see, for
example, Binney \& Tremaine 1987) can be important for halos of
$\lesssim 30$ particles. For such halos the evaporation time-scale,
especially in the presence of strong tidal fields, can be comparable
to or less than the Hubble time.  However, for halos containing $\grtsim
50$ particles, evaporation should be negligible. Nevertheless, the
trend with resolution seen in Table~\ref{xiparam} does suggest that
two-body evaporation is the process responsible for the differences.

The most likely explanation of these results is, in our opinion, the
accuracy of the time integration. The estimates of the evaporation
time-scale are done assuming no errors in energy exchange between
particles in a scattering event. Our analysis of such events in our
simulations, on the other hand, shows that with the time step of the
\ap3m~2-4 runs, there is severe energy conservation violation during
scatterings. The particles attain much larger velocities than they
should have if the integration were perfect. For example, in the
scattering event shown in Fig.~\ref{ipartlocal} the final kinetic
energy of the two particles is 100 times larger than their initial
kinetic energy. Therefore, a time step that is insufficiently small to
properly handle two-body scattering may exacerbate the process of
evaporation and result in much shorter than predicted evaporation
time-scales, even for halos containing relatively large numbers of
particles. The differences between AP$^3$M1 run and runs AP$^3$M2 and
4 indicate that this effect is very sensitive to both the force
resolution and to the time step of the simulation.

Another possible explanation is that halos do not evaporate but the 
particles are heated due to non-conservation of energy which makes 
halos ``puffier''. Such halos would be more prone to destruction 
by tides in high-density regions.

\subsection{Mass Function}

\begin{figure}
      \resizebox{\hsize}{!}{\includegraphics{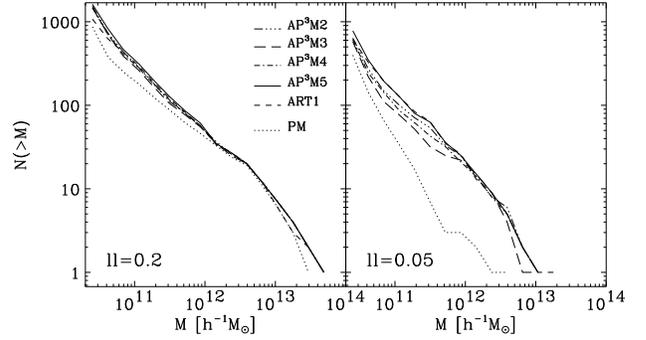}}
      \caption{Mass function for some simulations and 
               two linking lengths.}
      \label{massfunc}
    \end{figure}

In Fig.~\ref{massfunc} we show the mass functions of all halos
identified with various linking lengths at $z=0$. 

As can be seen in the left panel of Fig.~\ref{massfunc}, the halo mass
function in the PM run is biased towards large masses: the number of
objects of mass $\grtsim 2\times 10^{12}\hMsun$ in the PM run agrees
well with the corresponding number in the high-resolution runs, while
at lower masses the number of halos is strongly underestimated. Mass
$2\times 10^{12}\hMsun$ corresponds to the virial radius (at $z=0$,
$\delta_{\rm vir}=178$) of $R_{\rm vir}\approx 213\hkpc$, i.e. very
close to the force resolution of the PM runs ($234\hkpc$). The
conclusion is that the PM runs fail to reproduce the correct
abundances of halos with the virial radius less than about two grid
cell sizes.  

At smaller linking length, $ll=0.05$, (right panel of
Fig.~\ref{massfunc}) the PM run severely underestimates the halo
abundances. We attribute this also to the poor force resolution of the
run. The poor resolution prevents the formation of dense cores in the
inner regions of collapsed halos and halos that do not reach overdensity
of $\approx 11,000$ (overdensity of objects identified with $ll=0.05$)
will be missed. Even if the central density of some halos reaches this 
value, the halos are still considerably less dense than their counterparts
in the high-resolution runs and are therefore more susceptible to
destruction by the tidal fields in high-density regions (Moore et
al. 1996; Klypin et al. 1999).

   \begin{figure*}
      \resizebox{\hsize}{!}{\includegraphics{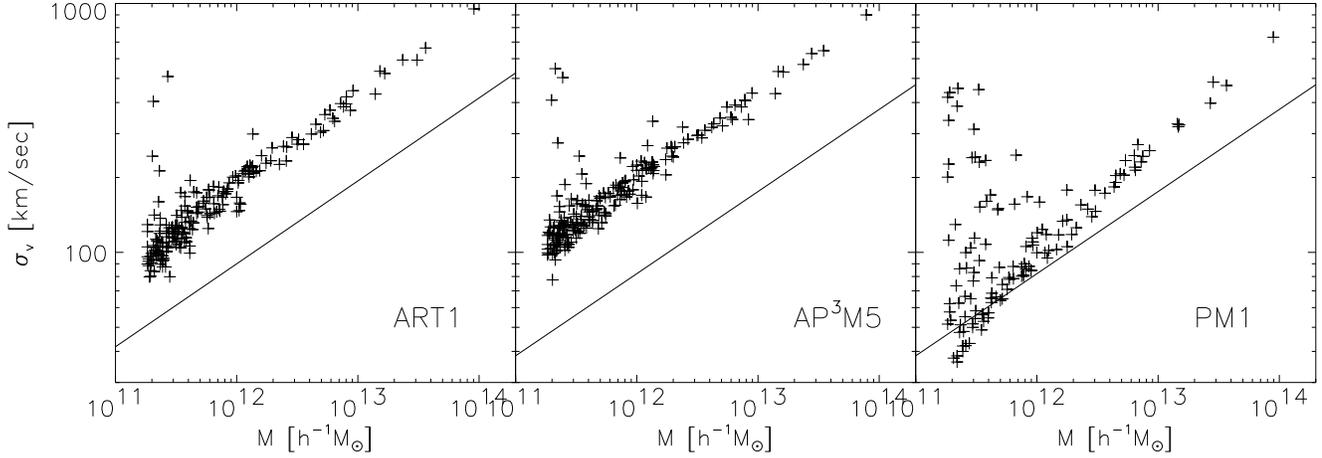}}
      \caption{Correlation between the velocity dispersion and
               mass of halos for
               ART1, AP$^3$M1, and PM. Only halos with more than
               50 particles are taken into account. The solid line
               denotes $\sigma_{\rm v} \propto M^{1/3}$}
      \label{mvdisp}
    \end{figure*}

The mass functions of the high-resolution runs agree to $\sim 30\%$
for isolated halos of overdensity $\delta = 178$ ($ll = 0.2$). The
\ap3m runs have a larger number of identified halos at masses $\lesssim
5\times 10^{10}\hMsun$ (i.e., halos containg $\lesssim 15$ particles)
than the ART runs.  This difference is due to smaller number of small
halos in low-density regions in the ART runs discussed in \S~3.2.  The
mass functions of the AP$^3$M1, AP$^3$M5, and the ART runs agree well
at masses $\grtsim 10^{11}\hMsun$ for both $ll=0.2$ and $ll=0.05$
(overdensity of 178 and 13000, respectively), indicating that both
runs produced similar populations of halos with similar central
densities. The mass functions of the \ap3m~2-4 runs are similar for
$ll=0.2$, but show differences for $ll=0.05$. Thus, for example, the
abundance of halos of mass $\sim 10^{11}-10^{12}\hMsun$ ($\sim 30-300$
particles) in the AP$^3$M3 run is underestimated by a factor of
$\approx 1.5-2$. The mass functions of the AP$^3$M2 and AP$^3$M4
runs lie in between those of the AP$^3$M3 and AP$^3$M5 runs.

The fact that differences are present at small linking length
indicates differences in the high-density regions. This may be due to
the generally lower inner densities of halos and/or to the destruction
of ``heated'' sattelites discussed above.

\subsection{Halo correlation function}

   \begin{figure}
      \resizebox{\hsize}{!}{\includegraphics{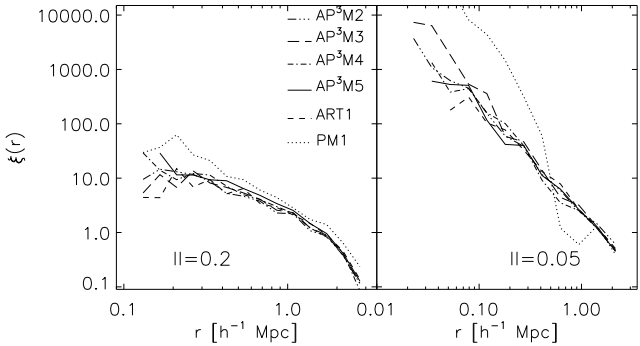}}
      \caption{Correlation function of halos for some simulations and 
               two linking lengths. A mass cut of 25 
               particles was used to construct halo catalogs.}
      \label{xi}
    \end{figure}

Figure~\ref{xi} shows the 2-point correlation function of identified
DM halos.  There is good agreement between the correlation functions of
isolated virialized ($ll=0.2$) halos in high-resolution
runs. Similar to the dark matter correlation function, the agreement
is better than 10\%.  The agreement between the AP$^3$M1, AP$^3$M5,
and the two ART runs does not break even at higher overdensities
($ll=0.05$), which indicates that these runs produced similar
small-scale substructures in the high-density regions within isolated
halos. We do not find any differences of the type seen in the DM
correlation function (\S~3.5) between \ap3m runs at $ll=0.2$. For
halos identified using $ll=0.05$ some differences are observed, but
these are comparable to the poisson errors and are therefore
inconclusive.

On the other hand, halos in the PM simulation exhibit higher
correlations than halos in the high-resolution runs. As discussed in
the previous section, the halo mass function in the PM run is biased
toward high masses. The higher amplitude of the correlation function
can thus be explained by the mass-dependent bias: higher mass
halos are clustered more strongly.

\subsection{Velocity Dispersion vs. Mass}

Figure~\ref{mvdisp} shows the correlation of velocity dispersion and
mass for one of the ART, \ap3m, and PM simulations. A correlation
$\sigma_{\rm v} \propto M^{1/3}$ is expected for virialized halos.
For the ART1 and AP$^3$M5 simulations, the best fit slope of the
correlation for the 50 most massive halos is 0.33 (the correlation for
the rest of the \ap3m runs is similar). The velocity dispersion of
low-mass halos scatters around this value.  For comparison, a line of
slope 1/3 is included in all three panels.  It should be mentioned
here that at the low mass end the FOF groups contains only a few
particles so that $\sigma_{\rm v}$ is not well determined because the
error due to unbound particles accidentally linked by the FOF algorithm
may be very high.

The halos in the PM run deviate from the expected slope at masses
$\lesssim 10^{13}\hMsun$. The virial radii of the halos of these
masses are $\lesssim 364\hkpc$. Their size is thus $\lesssim 3$ force
resolutions across. Therefore, the potential and the internal dynamics
of the particles in these halos are underestimated by the PM code
leading to a steeper slope of the $\sigma_{\rm v}- M$ relation.

\subsection{Density Profiles}

  \begin{figure*}
     \resizebox{\hsize}{!}{\includegraphics{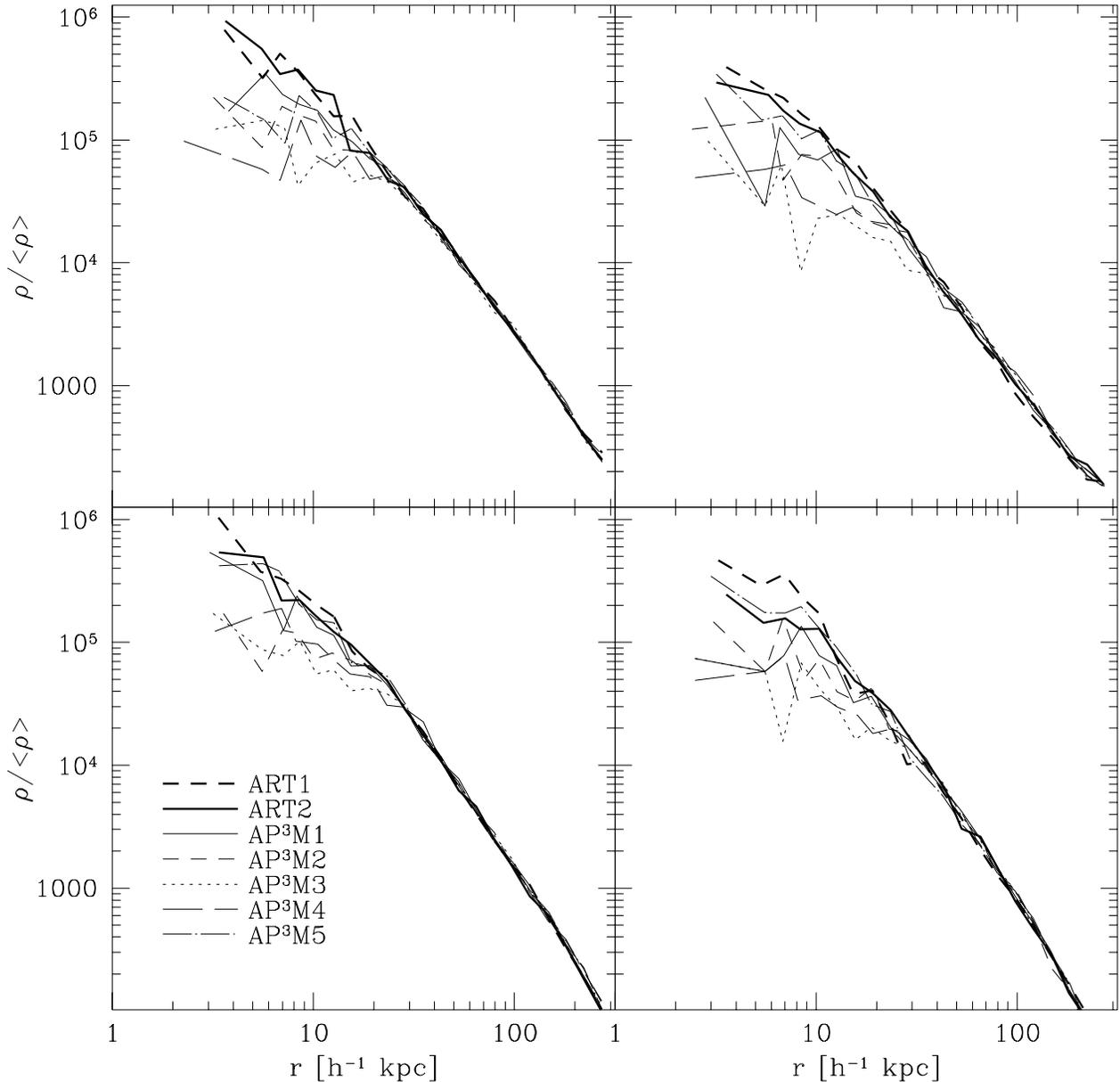}}
      \caption{Density profiles of halos in different runs.}
      \label{hpc}
   \end{figure*}

  \begin{figure*}
     \resizebox{\hsize}{!}{\includegraphics{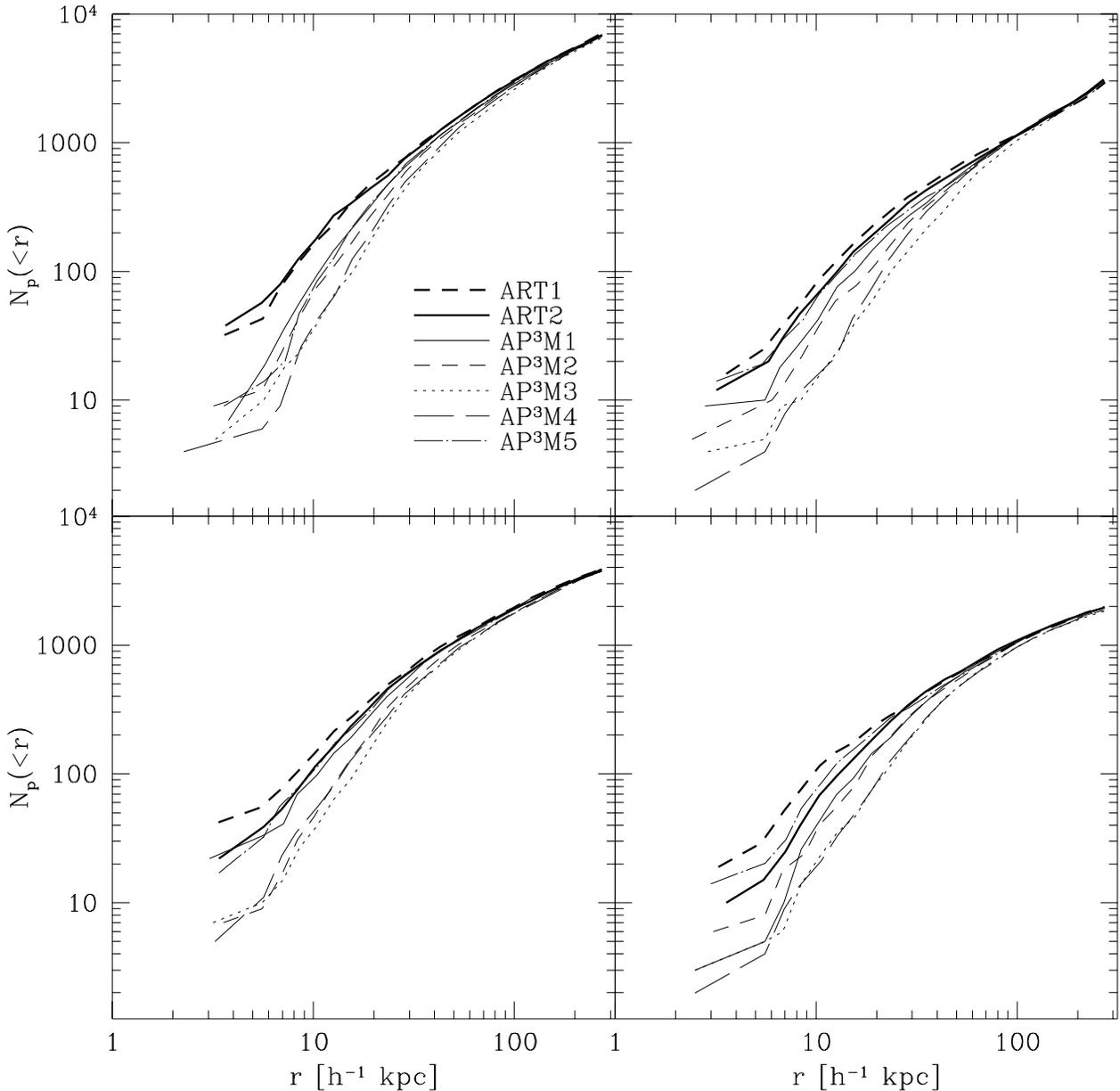}}
      \caption{Cumulative number of particles within radius $r$ from the centers
       of four halos shown in the previous figure.}
      \label{npc}
   \end{figure*}

In this and the next two sections we present halo-to-halo comparison
of individual halos in different simulations. In this section we will
compare the density profiles of DM halos.  The density distribution of
hierarchically formed halos is currently a subject of debate (see,
e.g., Navarro, Frenk \& White 1997; Kravtsov et al. 1998; Moore et
al. 1999) and study of the resolution effects and cross-code
comparisons are therefore very important.  In figure~\ref{hpc} we
present the density profiles of four of the most massive halos in our
simulations. We have not shown the profile of the most massive halo
because it appears to have undergone a recent major merger and is not
very relaxed. In this figure, we present only profiles of halos in the
high-resolution runs. Not surprisingly, the inner density of the PM
halos is much smaller than in the high-resolution runs and their
profiles deviate strongly from the profiles of high-resolution halos
at the scales shown in Fig.~\ref{hpc}. We do not show the PM profiles
for clarity.

A glance at Fig.~\ref{hpc} shows that all profiles agree well at
$r\grtsim 30\hkpc$. This scales is about eight times smaller than the
mean interparticle separation. Thus, despite the very different
resolutions, time steps, and numerical techniques used for the
simulations, the convergence is observed at a scale much lower than
the mean interparticle separation, argued by Splinter et al. (1998) to
be the smallest trustworthy scale.  At smaller scales the profiles
become more noisy due to poorer particle statistics (see
Fig.~\ref{npc}).

Nevertheless, there are systematic differences between the runs.  The
profiles in two ART runs are identical within the errors indicating
convergence (we have run an additional simulation with time steps
twice smaller than those in the ART1 finding no difference in the
density profiles). Among the \ap3m runs, the profiles of the AP$^3$M1
and AP$^3$M5 are closer to the density profiles of the ART halos than
the rest. The AP$^3$M2, AP$^3$M3, and AP$^3$M4, despite the higher
force resolution, exhibit lower densities in the halo cores, the
AP$^3$M3 and AP$^3$M4 runs being the most deviant. These differences
can be seen more clearly in Fig.~\ref{npc}, where we plot the
cumulative number of particles (i.e., mass) within radius $r$ for the
halos shown in Fig.~\ref{hpc}. The differences between AP$^3$M3 and
AP$^3$M4 and the rest of the runs are apparent up to the radii
containing $\sim 1000$ particles.

\begin{table}
\caption{The ratio of number of time steps to the dynamic range
         of simulations for the high-resolution runs. The number of time steps
         for the ART runs correspond to the time step on the highest refinement
         level.}
\label{nstep2dr}
 \begin{center}
 \begin{tabular}{|l||c|c|c|} \hline
 Run                      &  $N_{step}/{\rm dyn. range}$  \\ \hline 
 \raisebox{4mm}{}AP$^3$M4 &  0.469  \\ \hline
 \raisebox{4mm}{}AP$^3$M3 &  0.702  \\ \hline
 \raisebox{4mm}{}AP$^3$M2 &  0.938  \\ \hline
 \raisebox{4mm}{}AP$^3$M1 &  1.875  \\ \hline
 \raisebox{4mm}{}AP$^3$M5 &  3.750  \\ \hline
 \raisebox{4mm}{}ART2     &  2.578  \\ \hline
 \raisebox{4mm}{}ART2     &  5.156  \\ \hline
 \end{tabular}
 \end{center}
\end{table}

These results can be interpreted, if we examine the trend of the
central density as a function of the ratio of the number of time steps
to the dynamic range of the simulations (see Table~\ref{param}) shown
in Table~\ref{nstep2dr}. The ratio is smaller when either the number
of steps is smaller or the force resolution is higher.
Table~\ref{nstep2dr} shows that agreement in density profiles is
observed when this ratio is $\grtsim 2$. This suggests that for a
fixed number of time steps, there should be a limit on the force
resolution. Conversely, for a given force resolution, there is a lower
limit on the required number of time steps. The exact requirements
would probably depend on the code type and the integration scheme.
For the \ap3m code our results suggest that the ratio of the number of time
steps to the dynamic range should be no less than one. It is interesting
that the deviations in the density profiles are similar to and are
observed at the same scales as the deviations in the DM correlation
function (Fig.~\ref{DMxi}) suggesting that the correlation function is
sensitive to the central density distribution of dark matter halos.

These results are indicative of the sensitivity of the central density
distribution in halos to parameters of the numerical simulation. However,
due to the limited mass resolution of our test runs, they do not shed
light on the density profile debates. The profiles of the ART halos agree
well with those of the \ap3m halos, if the latter are simulated with a
sufficiently large number of time steps. But debated differences are
at scales of $\lesssim 0.01R_{\rm vir}$, which are not resolved in
these simulations. We are currently carrying out a more detailed,
higher-resolution study to clarify the issue.

\subsection{Shared Particles}

We have argued above that particle trajectories may diverge in
high-density regions due to their instability to small integration
errors. However, despite the instability of its trajectory, we can
expect that a bound particle will stay bound to its parent halo. To
this end we compare the particle content of individual halos in
different runs.  We have identified the counterpart halos in different
runs as systems that have the largest number of common (or shared)
particles.  This method is superior to coordinate based methods
because due to the phase errors (see \S~3.3) one does not expect to
find the halos at the exact same position in different runs (this is
especially not for small groups of particles).

   \begin{figure}
      \resizebox{\hsize}{!}{\includegraphics{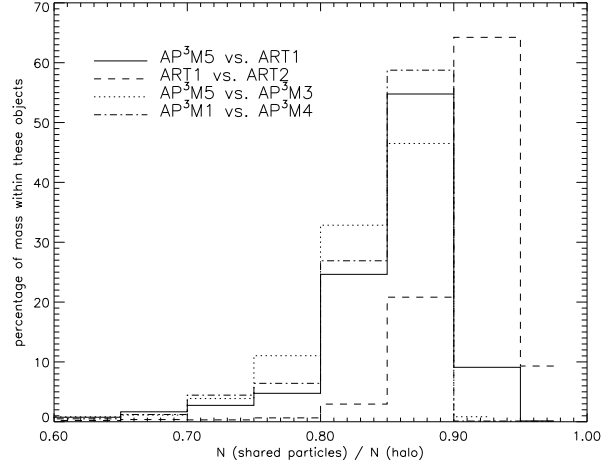}}
      \caption{Percentage of mass included in halos
               with a specific amount of shared particles and
               containing at least 25 particles.}
      \label{corrmass}
    \end{figure}

   \begin{figure}
      \resizebox{\hsize}{!}{\includegraphics{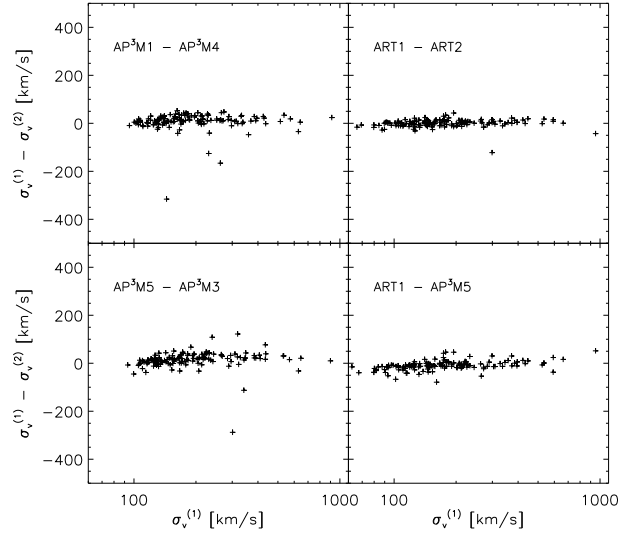}}
      \caption{Correlation of the velocity dispersions of the same
                halos identified in two different simulations. Only 
                halos with more than
                50 particles are taken into account}
      \label{corrvdisp}
    \end{figure}

Fig.~\ref{corrmass} shows the percentage of mass that can be found in
particle groups that coincide in a specific amount of shared
particles. The figure shows that in all runs most halos have more than
$80\%$ of their particles in common. The distributions peak at around
85-90\% for all of the comparisons, except ART1 vs. ART2 which peaks
at $\approx 90-95\%$.  Even though a direct comparison of particle
positions within halos is not a very useful way of comparing different
runs, this result shows that the particle content of halos is rather
similar.

\subsection{Correlation of Velocity Dispersions}

In Fig.~\ref{corrvdisp} we compare the velocity dispersions of halos
identified in different simulations by the FOF algorithm assuming a
minimum particle number of 50 per halo. The figure shows that the
velocity dispersions agree reasonably well with the overall scatter of
$\sim 50\kms$. The very few outliers are halos of very different
masses identified as the same halo by the FOF algorithm. Small
differences in particle distribution may result in the identification
of a binary system in one simulation and only a single halo in another
simulation, leading to the large differences in velocity dispersions.

The differences between velocity dispersions appear to be independent
of halo mass, and although they reach $\sim 50\%$ for the lowest mass
halos, there are no obvious systematic differences between different
simulations.

\section{Discussion and conclusions}

We have presented results of a study of resolution effects in
dissipationless simulations. As we noted in the introduction, an
additional goal of this study was to compare simulations done using
two different high-resolution $N$-body codes: the \ap3m and the
ART. Our results indicate that both codes produce very similar results
at all scales resolved in the presented simulations, given the force
resolution and time step are such that convergence is reached within
the code type. Our results indicate that numerical effects may be
complicated due to a combined effect of mass and force resolution and
inaccuracies of the time integration scheme. The precise magnitude of
the effects depends on the numerical parameters used and the considered 
statistics or property of particle distribution.

Particles in dissipationless cosmological simulations are supposed to
represent elements of the dark matter distribution in phase space
\footnote{Unless
particles are supposed to represent individual galaxies, which is
virtually never the case in modern simulations.}. The
smaller the particle number, the larger the volume associated
with each of the particles.  During the course of evolution, according
to Liouville's theorem, the phase-space volume of each element
should be preserved. Its shape, however, will be changing.
Correspondingly, the eulerian space volume of these elements may
shrink (for particles in an overdense collapsing region of space)
or expand (for particles in underdense regions). Regardless of its
initial shape, each element can be stretched due to the anisotropic
nature of the gravitational collapse in cosmological models.

Usually, none of these effects is modelled in cosmological $N$-body
simulations. The gravitational field of each particle is assumed to be
roughly isotropic and therefore the effects of the volume stretch
cannot be modelled. These effects are not addressed in the present
study, because they can only be studied by widely varying the number
of particles at a fixed force resolution and for a fixed time
step\footnote{Due to the phase errors discussed below, such a study
should also be carried out using the same numerical code.}, while we
have varied the force resolution and time step keeping the number of
particles fixed. A large number of particles corresponds to the smaller
phase-space associated with each particle and the symmetric particle
approximation is more accurate. Convergence studies of the halo
density profiles indicate that these effects are small (at least at
radii $\grtsim 0.02-0.05R_{\rm vir}$).  However, this does not mean
that they cannot be important for other statistics.

Nevertheless, we can study other kinds of resolution
effects. Normally, softening of interparticle gravitational force
should be approximately equal to the spatial size of the phase-space
volume element associated with each particle. If the softening is much
smaller than this size, the volume elements will behave like particles
and two-body scattering is possible.  This was indeed observed in some
of our high-resolution simulations. This scattering contradicts the
collisionless nature of the modelled dark matter and is thus
undesirable.

For the mass resolution of our simulations ($\approx 3.55\times
10^9\hMsun$), scattering was observed in simulations with uniform
force resolution of $\lesssim 3\hkpc$ and disappears for larger values
of resolution. While we find strong scattering for $\approx 1.4\%$ of
the particles, many more may have suffered weaker scattering events
during the course of simulation. Indeed, we find indirect evidence
that scattering has substantial effect on the particle
distribution. In particular, we find that it may noticeably affects
the small-scale amplitude of the 2-point correlation function of dark
matter, the abundance and mass function of dark matter halos, and the
central density distribution of halos. The effect is amplified
strongly for simulations in which larger time steps are used.  This is
because for larger time steps the scattering events severely violate
energy conservation. The magnitude of the violation is very sensitive
to the time step: the amount of momentum gained by a particle during a
scattering event is $\propto g\Delta t$, where $g$ is the
acceleration.  Therefore, for the same force resolution (which means
that the same $g$ can be achieved), the momentum gain will be
proportional to $\Delta t$.

The 2-point correlation function of matter and the central density
distribution in halos in our runs are affected by these effects at
scales as large as $\approx 15-20$ force softenings.  Also, the
abundance of satellite halos in high-density regions appears to
be affected as well.  We think that these effects may be due to the
following two phenomena. First, if the time step of the simulation is
too large for a given force resolution, particle trajectories are not
integrated accurately in the highest density regions, where the
gradient of potential is the highest. In some sense, particles scatter
off the central density peak, and may gain energy when integrated with
large time step (see arguments above). 
The number of simulation time steps per deflection on a scale $R$ can be estimates as
\begin{equation}
N_{\rm step}^{\rm }= \frac{R}{v_{100}\Delta t}
\approx 7.51\times 10^{-4}R N_{\rm step}^{\rm tot}v_{100}^{-3},
\end{equation}
where $R$ is in kpc, $\Delta t$ and $N_{\rm step}^{\rm tot}$ is the
time step and the total number of time steps of the simulations, and
$v_{100}$ is particle velocity in units of $100\kms$. We have assumed
the Einstein-de Sitter universe with the Hubble constant of
$H_0=50\kms{\ }{\rm Mpc^{-1}}$. For high-velocity particles streaming
through the centers of massive halos, there may be just a few time
steps to integrate the part of trajectory where strong changes in
acceleration occur.  As illustrated in Fig.~\ref{ipartlocal}, this may
not be sufficient to ensure energy preservation and may result in an
energy gain by particles.  This leads to artificial heating and lowers
the central density because having acquired energy, particles are
not as likely to enter the central region of halo.

The second phenomenon is due to the ``graininess'' of the potential.
Particles in the high-density regions may feel the discreteness of the
density field and suffer scattering. We do find evidence for
scattering in the high-density regions in our simulations (see
\S~3.4).  Here, again, the effect may be amplified strongly by
the incorrect integration of such scattering. Indeed, run AP$^3$M1
($N_{\rm step}^{\rm tot}=8000$) performs much better than the run
AP$^3$M4 ($N_{\rm step}^{\rm tot}=2000$), although both runs have the
same mass and force resolution.

The two effects described above may operate in combination, although the
first effect does not depend on the mass resolution. The second effect
should be eliminated for higher mass resolution. Both lead to the
artificial heating of particles thereby lowering the central density
of halos and possibly ejecting the particles altogether in some
cases. Visual comparisons of halos in the AP$^3$M3 (the run which
performed the worst) and the AP$^3$M5 runs shows that AP$^3$M3 halos
appear ``puffier'' and more extended than the same halos in the
AP$^3$M5 or ART runs. Puffier halos may be destroyed more easily by
tides in high-density regions which may explain some of the
differences seen between the mass functions of halos in different runs.

Our results show that in constant time step high-resolution
simulations the total number of time steps must be rather high to
ensure good energy conservation. This requirement can become
computationally prohibitive in simulations that follow large numbers
of particles. In case of the {\ap3m} code, it would probably be
preferrable to use its version in the publicly available code
``Hydra'' (Couchman et al. 1995) that uses adaptively varied time
step.

The conditions for scattering, discussed in \S~3.4, occur if the force
softening is smaller than the scale $s$, which {\em in units
of mean interparticle separation\/} is
\begin{equation}
\tilde{s}\approx 1.209\times 10^{-3}\Omega_0^{1/3}\left(\frac{v}{100\kms}\right)^{-2}
\left(\frac{m_p}{10^8\hMsun}\right)^{2/3},
\end{equation}
{\em and\/} is considerably smaller than the {\em local\/} interparticle
separation: $d_{\rm loc}=(1+\delta)^{-1/3}$ (in units of the mean
interparticle separation, $\delta$ is the local particle
overdensity). For our simulations $s\approx 3v^{-2}_{100}\hkpc$,
where $v_{100}$ is particle velocity in units of
$100\kms$. This means that condition $\tilde{s}\ll d_{\rm loc}$ is satisfied
everywhere but in the highest density regions: $\delta\grtsim
10^4$. The conditions for strong scattering occur for the slow moving
($\lesssim 100\kms$) particles in the \ap3m runs $1-4$. Such slow
moving particles are likely to be present in the low-density regions
and in small-mass halos of velocity dispersion $\sigma_{\rm v}\lesssim
100-200\kms$. This may explain our result that halos of mass 
$\lesssim (0.5-1)\times 10^{12}\hMsun$ ($\sigma_{\rm v}\lesssim 100-150\kms$, 
see Fig.~\ref{mvdisp}) appear to be affected by scattering. 

One may question the relevance of these results, given the small size
of the simulations and extremely high force resolution. Note, however,
that our results would be applicable (save for the presence of very
massive clusters) to any $256^3$-particle simulation, in which force
resolution of considerably smaller than the scale $\tilde{s}$ is
adopted.  The simulations with the particle mass and dynamic range not
very far from ours, have already been done. For example, all
$256^3$-particle simulations of $239.5\hMpc$ box presented by Jenkins
et al. (1998) satisfy the condition of $\epsilon < \tilde{s}$ (where
$\tilde{s}$ is estimated using the above equation for $v_{100}\lesssim
1.5$) and have been run using $<1600$ time steps. The parameters of
the recent ``Hubble volume'' simulation (Colberg et al. 1998) also
satisfy conditions for strong scattering: $s\approx 1.9\hMpc$, while
force resolution of the simulations is $\epsilon=100\hkpc$.  The
particle mass of these simulations is $\approx 2\times 10^{12}\hMsun$,
which means that individual particles represent galaxies rather than a
phase-space element. Galaxies form a collisional system so the
presence of scattering may be considered as a correct model of the
evolution of galaxy clustering. Moreover, the mean interparticle
separation in these simulations is $\approx 2-3\hMpc$ and thus
conditions for scattering may only occur in the underdense regions.

A high-resolution $256^3$-particle run was also done recently using
the ART code (see, for example, Col\'{\i}n et al. 1999). However, for
this simulation ($m_p=1.1\times 10^9\hMsun$; $\Omega_0=0.3$) the scale
of strong scattering is $s\approx 0.94 v_{100}^{-2}\hkpc$, while
peak resolution is $\approx 4\hkpc$. The time steps for the particles
at the refinement level of this resolution corresponds to effective
number of time steps of $\approx 41,000$. Therefore, for this
simulation the strong scattering condition is not satisfied. Moreover,
a refinement level $L$ is introduced in these simulations only if the
local overdensity is higher than $\delta=5\times 2^{3(L+1)}$, or for
the highest resolution level $L=6$: $\delta\approx 10^7$. For these
overdensities, the local interparticle separation is $\approx 1\hkpc$,
and two-body interactions are thus unlikely.

To summarize, scattering can be precluded if the choice of force
resolution is guided by the scale $s$, which, in turn, depends on the
mass resolution (particle mass). This conclusion may seem 
similar to that of Splinter et al. (1998), who concluded that force
resolution should not be smaller than the mean interparticle
separation.  It is, however, quite different in practice: eq.~\eq_s2
shows that for our box size $s\approx 1$ in units of mean
interparticle separation only for $m_p\approx 3\times
10^{12}\hMsun$. For our mass resolution, force softening as small as
$5-10\hkpc$ is justified. This is $\sim 25-50$ times below the mean
interparticle separation.

We think that the conclusion of Splinter et al. is (at least in part)
due to the interpretation of poor cross-correlation between different
simulations on small scales as erroneous evolution in high-resolution
runs. Our analysis, presented in \S~3, shows that poor
cross-correlation is due to phase errors whose major source is
cumulative errors due to the inaccuracies of time integration. The
trajectories of particles become chaotic in high-density regions and
small differences in time integration errors tend to grow quickly. 

For this reason it may prove to be very difficult to get rid of this
effect by improving the time integration. Therefore, one should keep 
these errors in mind if a phase sensitive statistic is analyzed. 
Luckily, most of the commonly used statistics are phase-insensitive
and are not affected by such errors. Moreover, the errors are confined
to the small-scale high-density regions, and no significant 
phase errors are present in our simulations if the density field is smoothed 
on a scale $\grtsim 1\hMpc$. 

While this is clearly still an error, it has nothing to do with the
mass or force resolution and would be present even if both were
perfect. This point is clearly demonstrated by the fact that
simulations run using two different implementations of the PM code
correlate perfectly within the code type but cross-correlate rather
poorly when cross-code comparisons are made (see \S~3.3). Note that in
all of these PM runs, the force resolution is approximately equal to
the mean interparticle separation. 

The main conclusion of our study is that care must be taken in the
choice of force resolution for simulations. If a code with spatially
uniform force resolution is used, conditions for strong two-body
scattering may exist if the force resolution is smaller than the scale
$s$ discussed above.  The presence of scattering itself may not be
important (albeit undesirable); the relaxation time for systems, for
example, may be much longer than the Hubble time (e.g., Hernquist \&
Barnes 1990; Huang, Dubinski \& Carlberg 1993).  Its effects, however,
may be greatly amplified if the time step of the simulation is not
sufficiently small. In this case, severe violation of energy
conservation occurs during each scattering which may lead to
artificial injection of energy into the system.

\section*{Acknowledgements}

AVK and AAK are grateful to the Astrophysikalishes Institut Potsdam
(AIP), where this project was initiated, for the hospitality during
their visit.  We thank referee for useful comments.  This work was
funded by NSF and NASA grants to NMSU.  SG acknowledges support from
Deutsche Akademie der Naturforscher Leopoldina with means of the
Bundesministerium f\"ur Bildung und Forschung grant LPD 1996.  Our
collaboration has been supported by the NATO grant CRG 972148.


\end{document}